\documentclass[12pt]{article}
\usepackage[a4paper]{geometry}
\usepackage[italian,english]{babel}
\usepackage[T1]{fontenc}

\usepackage{amsmath, accents}
\usepackage{amsfonts}
\usepackage{amstext}
\usepackage{amssymb}
\usepackage{amsthm}
\usepackage{amscd}
\usepackage{mathrsfs}
\usepackage{bbold}
\usepackage{dsfont}
\usepackage{bbm}

\usepackage[pagebackref,draft=false]{hyperref}
\hypersetup{colorlinks,
linkcolor=myrefcolor,
citecolor=mycitecolor,
urlcolor=myurlcolor}

\usepackage[capitalize]{cleveref}
\usepackage{cite}
\usepackage{caption}
\usepackage{etaremune}

\usepackage{xcolor}
\definecolor{myurlcolor}{rgb}{0,0,0.4}
\definecolor{mycitecolor}{rgb}{0,0.5,0}
\definecolor{myrefcolor}{rgb}{0.5,0,0}
\usepackage{graphicx}
\usepackage{tikz}
\usepackage{tikz-cd}
\usetikzlibrary{graphs,decorations.pathmorphing,decorations.markings}

\usepackage{lipsum}
\usepackage{etoolbox}
\usepackage{makeidx}
\usepackage{sectsty}
\usepackage{dsfont}
\usepackage{enumitem} 
\usepackage[]{latexsym}
\usepackage{braket}
\usepackage{caption}
\usepackage[utf8]{inputenx}
\usepackage{lmodern}
\usepackage{textcomp}
\usepackage{microtype}
\usepackage{totcount}
\usepackage{blindtext}
\usepackage{ulem}

\newtheorem{theorem}{Theorem}[section]

\newtheorem{proposition}[theorem]{Proposition}

\newtheorem*{proof*}{Proof}

\newcommand{\be}{\begin{equation}}
\newcommand{\ee}{\end{equation}}
\newcommand{\bea}{\begin{eqnarray}}
\newcommand{\eea}{\end{eqnarray}}
\newcommand{\vsp}{\vspace{0.4cm}}

\newcommand{\blue}[1]{\textcolor{blue}{{#1}}}
\newcommand{\red}[1]{\textcolor{red}{#1}} 

\newcommand{\ac}{\mathscr{S}}
\newcommand{\el}{\mathbb{EL}}
\newcommand{\m}{\mathscr{M}}
\newcommand{\mm}{\mathbb{M}}
\newcommand{\pe}{\mathcal{P}(\mathbb{E})}

\newcommand{\fpe}{\mathcal{F}_{\mathcal{P}(\mathbb{E})}}
\newcommand{\tfpe}{\mathbf{T}_\chi \mathcal{F}_{\mathcal{P}(\mathbb{E})}}

\newcommand{\pssos}{\Pi^\star_\Sigma \Omega^\Sigma}
\newcommand{\os}{\Omega^\Sigma}

\newcommand{\pss}{\Pi^\star_\Sigma}

\newcommand{\upe}{U_{\mathcal{P}(\mathbb{E})}}

\newcommand{\ltre}{x^2 \partial_1 - x^1 \partial_2}
\newcommand{\ldue}{x^1 \partial_3 - x^3 \partial_1}
\newcommand{\luno}{x^3 \partial_2 - x^2 \partial_3}

\newcommand{\elag}{\mathcal{E}\mathscr{L}}

\newcommand{\dd}{{\rm d}}
\newcommand{\de}{\partial}

\title{Symmetries and covariant Poisson brackets \\ 
on pre-symplectic manifolds}

\author{F. M. Ciaglia$^{1,6}$ \href{https://orcid.org/0000-0002-8987-1181}{\includegraphics[scale=0.7]{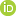}}, F. Di Cosmo$^{1,2,7}$ \href{https://orcid.org/0000-0003-0256-5913}{\includegraphics[scale=0.7]{ORCID.png}}, A. Ibort$^{1,2,8}$ \href{https://orcid.org/0000-0002-0580-5858}{\includegraphics[scale=0.7]{ORCID.png}}, \\ G. Marmo$^{3,4,9}$ \href{https://orcid.org/0000-0003-2662-2193}{\includegraphics[scale=0.7]{ORCID.png}}, L. Schiavone$^{1,5,10}$  \href{https://orcid.org/0000-0002-1817-5752}{\includegraphics[scale=0.7]{ORCID.png}}, A. Zampini$^{3,5,11}$ \href{https://orcid.org/0000-0003-0980-6003}{\includegraphics[scale=0.7]{ORCID.png}} \\
\footnotesize{$^{1}$\textit{ Depto. de Matem\'aticas, Univ. Carlos III de Madrid, Legan\'es, Madrid, Spain}} \\
\footnotesize{$^{2}$\textit{ ICMAT, Instituto de Ciencias Matem\'{a}ticas (CSIC-UAM-UC3M-UCM)}} \\
\footnotesize{$^{3}$\textit{ INFN-Sezione di Napoli, Naples, Italy}} \\
\footnotesize{$^{4}$\textit{ Dipartimento di Fisica ``E. Pancini'', Universit\`a di Napoli Federico II,  Naples, Italy}} \\
\footnotesize{$^{5}$\textit{ Dipartimento di Matematica e Applicazioni "Renato Caccioppoli", Università di Napoli Federico II, Napoli, Italy}} \\
\footnotesize{$^{6}$\textit{e-mail:\texttt{fciaglia[at]math.uc3m.es}}}, \footnotesize{$^{7}$\textit{e-mail:\texttt{fcosmo[at]math.uc3m.es}}},\\
\footnotesize{$^{8}$\textit{e-mail:\texttt{albertoi[at]math.uc3m.es}}}, 
\footnotesize{$^{9}$\textit{e-mail:\texttt{marmo[at]na.infn.it}}},\\
\footnotesize{$^{10}$\textit{e-mail:\texttt{luca.schiavone[at]unina.it}}}, 
\footnotesize{$^{11}$\textit{e-mail:\texttt{azampini[at]na.infn.it}}} 
}

\begin{document}

\maketitle

\vskip 1cm

\hfill\begin{minipage}[b]{12cm}
\footnotesize\textit{\noindent We dedicate this paper to the memory of Alexander M. Vinogradov, \\ who has been for us a friend, a colleague, a teacher, a source of \\inspiration for his work on the geometry of partial differential equations.}
\end{minipage}\medskip

\begin{abstract}
Noticing that the space of the solutions of a first order Hamiltonian field theory has a pre-symplectic structure, we describe a class of conserved charges on it associated to the momentum map determined by any symmetry group of transformations. 
Gauge theories are dealt with by using a symplectic regularization based on an application of Gotay's coisotropic embedding theorem. The analysis of Electrodynamics and of the Klein-Gordon theory illustrates the main results of the theory as well as the emergence of the energy-momentum tensor algebra of conserved currents.
\end{abstract}
\bigskip

\section*{Introduction}
\addcontentsline{toc}{chapter}{Introduction}

Symmetries play a crucial role in the development and the analysis  of physical theories. 
Beyond the paradigmatic examples in classical field theory of Einstein's equation in general relativity or the Yang-Mills equation for gauge fields,  symmetries are used, for instance, to determine the form of the Lagrangian (or respectively of the Hamiltonian) function within the Lagrangian (respectively Hamiltonian) formulation of the theory (an interesting example in this perspective is given by Utiyama's theorem). 
Symmetries are also used to uncover significant global properties of the theory itself, like the existence of different phases or sectors, or even as an effective tool to analyse its quantum aspects (ranging from the study of the properties of quantum states, to the study of the renormalizability or to obtain definite predictions when anomalies arise).   

In most cases these symmetries of interest can be termed ``geometrical'' as they emerge from the underlying geometrical structures (consider for instance those symmetries coming from geometrical symmetries of the space-time of the theory).
This is the case of the relativistic covariance of a field  theory associated to the action of the Poincar\'e group on the (vector)  bundle used to formulate the theory itself (see, Sect. \ref{Sec:Symmetry theory on the Euler-Lagrange space}), or of the invariance under the action of a gauge group, described in terms of identity based automorphisms of a suitable (principal) bundle (see Sect. \ref{Subsec:Gauge theories: free Electrodynamics}). 
Other examples of geometrical symmetries include conformal invariance, instrumental in the analysis of quantum field theories in 2 dimensions or the group of causal diffeomorphisms of a space-time in the case of theories of gravitation.

There is another use of symmetries that runs parallel to the development of the previous physical
motivations. 
They are instrumental, from a mathematical point of view, to analyze the problem of the integrability of differential equations.
Indeed, they emerge from the discovery of integrable hierarchies of non-linear partial differential
equations like the KdV equation or the KP hierarchy. 
In such cases the integrability properties of the model are associated to rich geometrical data like a bi-Hamiltonian structure or a Lax representation
of the equation, that often can be interpreted as the existence of families of transformations on the space of the solutions of the equations. 
Examples of such symmetries are related to the Miura transform in the KdV case, to the Backlünd transformations, or to Darboux transformations for integrable systems obtained from a factorization property.

These aspects, and many more, have been extensively analysed in the last two centuries both within the mathematical and theoretical physics literature: 
a (far from exhaustive) list of contribution where the previous ideas have been developed is  \cite{Utiyama1956, Weyl1952-Symmetry, Wigner1959-Group_theory, Vinogradov1981-Geometry_Nonlinear_equations, Vinogradov-Krasil-Lychagin1986-Introduction_geom_diff, Weinberg1995-QFT1, Yvette2011-Noether, Sardanashvily2016-Noether, Marmo-Schiav-Zamp2020-NoetherI, Marmo-Schiav-Zamp2021-Noether2, Costa-Forger-Pegas2018-Groupoids_Noether, Costa-Forger-Romero2021-Groupoids_Gauge_theories, Gaset-Roman-Roy2021-Symmetries_multisymplectic}, where also Noether's theorem, that is, the relation between symmetries and the so called conservation laws is elucidated. 
In particular, in recent references where classical field theories are analysed within the setting of jet bundles and their duals, conserved currents associated with symmetries of an action functional are modelled as $(m-1)$-forms on a fibre bundle underlying the theory (with $m$  the dimension of the space-time on which the theory is developed) (see, for instance, \cite{Carinena-Crampin-Ibort1991-Multisymplectic, Krupka2015-Variational_Geometry, Sardanashvily2016-Noether, Asorey-Ibort-Spivak2017-Boundary_conditions}).

In the present paper  we show how  the above mentioned conserved currents emerge from the momentum map associated to a symmetry group action on the space of solutions of the theory.
In particular, we focus on first order Hamiltonian field theories whose setting is developed in \cite{Carinena-Crampin-Ibort1991-Multisymplectic, Ibort-Spivak2017-Covariant_Hamiltonian_YangMills}.

The space of solutions of first order Hamiltonian field theories turns to be a presymplectic manifold, which is locally diffeomorphic (within the multisymplectic formalism) to the space of integral curves of a (pre-)symplectic system constructed on a space of fields on an arbitrary slice of the space-time of the theory  (see Sect. \ref{Sec:The Euler-Lagrange space as pre-symplectic manifold}, Prop. \ref{prop:presymplectic}, and \cite{Ibort-Spivak2017-Covariant_Hamiltonian_YangMills, Asorey-Ibort-Spivak2017-Boundary_conditions, Ciaglia-DC-Ibort-Marmo-Schiav2020-Jacobi_Particles, Ciaglia-DC-Ibort-Marmo-Schiav2020-Jacobi_Fields}).
It turns out that a class of conserved charges, defined directly on the space of solutions of the theory, is defined in terms of the momentum map associated to a suitable symmetry group, extending the well known Hamiltonian version of Noether's theorem (see, for instance, \cite{Ortega-Ratiu2004-Momentum_map} for a comprehensive review) to a global (pre-)symplectic covariant description of field theories. 

Gauge theories are identified with theories such that the space of solutions have a pre-symplectic structure with non-trivial characteristic distribution.  
The conserved charges associated to groups of symmetry of the theory are studied on a symplectic formalism, i.e., restoring its natural algebraic structure by using the symplectic regularization of the theory provided by the coisotropic embedding theorem (see, for instance, \cite{Gotay-Sniatycki1981-Coisotropic_embeddings}).
We find it relevant to notice that one of the advantages of this approach is that it allows to consider within the same formalism the  so called Noether's first and second theorems, the difference determined by the intersection of the orbits of the group and the characteristic distribution of the pre-symplectic structure.   

The paper is organized as follows.
The first part of the paper (Sect. \ref{Sec:Symmetries for pre-symplectic dynamics}) is devoted to review the basic structures of the theory of canonical symmetry groups on symplectic and pre-symplectic manifolds respectively and, in the second case, their symplectic realisations by using the equivariant extension of the coisotropic embedding theorem.
Sect. \ref{Sec:The Euler-Lagrange space as pre-symplectic manifold} is devoted to succinctly review the geometry of the space of solutions of a first order Hamiltonian field theory and, in particular, its natural pre-symplectic structure, which is the arena where we construct a theory of symmetries.
In Sect. \ref{Sec:Symmetry theory on the Euler-Lagrange space}, we apply the abstract ideas developed in Sect. \ref{Sec:Symmetries for pre-symplectic dynamics} to the Euler-Lagrange space of a Hamiltonian field theory described in Sect. \ref{Sec:The Euler-Lagrange space as pre-symplectic manifold} by considering two significant examples: a symplectic system, the Klein-Gordon theory (Sect. \ref{Subsec:The symplectic case: Klein-Gordon theory}), and a gauge theory, free Electrodynamics (Sect. \ref{Subsec:Gauge theories: free Electrodynamics}).


\section{Symmetries for pre-symplectic dynamics}
\label{Sec:Symmetries for pre-symplectic dynamics}

We devote this section to review the geometrical theory of symmetries that can be constructed for dynamical systems admitting a symplectic or pre-symplectic formulation. 
We refer, for instance, to \cite{Ortega-Ratiu2004-Momentum_map, Marmo-Schiav-Zamp2020-NoetherI, Marmo-Schiav-Zamp2021-Noether2}, and references therein, for a more detailed exposition.

\paragraph{Pre-symplectic manifolds.}

A pre-symplectic manifold $(\mathcal{M},\, \omega)$ is a smooth manifold $\mathcal{M}$, equipped with a closed 2-form $\omega$, i.e. such that $\dd\omega=0$. 
As a manifold we mean a smooth not necessarily finite dimensional manifold. 
In the infinite dimensional case $\mathcal{M}$ will be assumed to be a paracompact, second countable, Hausdorff Banach manifold (see, for instance, \cite{Abraham-Marsden-Ratiu2007-Manifolds, Michor1980-Differentiable_mappings}, where both cases are analysed) so that a rigorous definition of the tangent $\mathbf{T}\mathcal M$ and the cotangent $\mathbf{T}^{\star}\mathcal M$ bundles over $\mathcal{M}$ exist, as well as  the standard Cartan  differential calculus.  
Associated to the 2-form $\omega$ it is immediate to define its characteristic set
\be
\label{uno}
\mathrm{ker}_m\, \omega \,=\, \left\{\, X_m \in \mathbf{T}_m \mathcal{M} \;\; : \;\; \omega_m(X_m, V_m) \,=\, 0 \;\; \forall \,\, V_m \in \mathbf{T}_m \mathcal{M} \,\right\}
\ee
at any point $m\in\mathcal M$. 
When this set is zero, $(\mathcal M, \omega)$ is said to be weakly symplectic; when $\omega$ induces an isomorphism between $\mathbf{T}_m \mathcal{M}$ and $\mathbf{T}^\star_m \mathcal{M}$, then $(\mathcal M, \omega)$ is said strongly symplectic. 
It is easy to see that such notions coincide when $\mathcal M$ is finite dimensional, or more generally when the Banach space modelling $\mathcal M$ is isomorphic to its Banach dual. 
When the characteristic set $\ker_m\omega$ is not zero and defines a  vector subbundle of $\mathbf{T}\mathcal{M}$, it will be called the characteristic bundle.   
We will denote it by $\mathbf{K} \to \mathcal{M}$ whereas we will denote by $K_m$ the distribution $\mathrm{ker}_m \omega$ on $\mathcal{M}$, which we will call the characteristic distribution.
Defined as the kernel of a closed 2-form, such characteristic  distribution turns to be involutive, i.e., if $X$ and $X'$ are vector fields in $K$, its Lie bracket $[X, X']$ is again in $K$.
Thus, the space of smooth sections of $\mathbf{K}$, say $\Gamma(\mathbf{K})$ is a Lie algebra (possibly infinite dimensional).
The space of leaves of the distribution $K$ gives rise to a foliation of $\mathcal{M}$.
A leaf $\mathcal{N}$ is a maximal connected integral submanifold of the distribution $K$.
The space of leaves, denoted by $\mathcal{M}/K$, carries a natural topological structure defined as the coarsest topology that makes the canonical projection
\be 
\mathrm{pr} \;\; : \;\; \mathcal{M} \to \mathcal{M}/K  \;\; :\;\; m \mapsto \mathcal{N}_m
\ee
continuous, where $\mathcal{N}_m$ is the leaf of $K$ passing through $m$.
In general $\mathcal{M}/K \,=:\, \mathcal{M}_\omega$ is not a smooth manifold, not even Hausdorff.
When this is the case and when the map $\mathrm{pr}$ is a submersion, then $\mathcal{M}_\omega$ inherits a canonical symplectic structure $\omega_K$ defined as $$\omega_K(\tilde{X}, \tilde{Y}) \,=\, \omega(X, Y)$$ where $\tilde{X},\, \tilde{Y} \in \mathbf{T}_{\mathcal{N}} \mathcal{M}_\omega$ and $X,\, Y \in \mathbf{T}_m \mathcal{M}$, with $\mathrm{pr}(m)\,=\, \mathcal{N}$ and $X,\, Y$ two arbitrary lifts of $\tilde{X},\, \tilde{Y}$ to $\mathcal M$.

\paragraph{Hamiltonian systems.}
A Hamiltonian system on a pre-symplectic manifold $(\mathcal{M},\, \omega)$ is defined as a triple $(\mathcal{M},\, \omega,\, H),$ where $H$ is a smooth real valued function on $\mathcal{M}$ called a Hamiltonian, with corresponding Hamiltonian  vector field $\Gamma$ on $\mathcal{M}$ satisfying
\be \label{Eq:Hamiltonian equation}
i_\Gamma \omega \,=\, \dd H \,.
\ee
When $\omega$ is not weakly symplectic,  the vector field $\Gamma$ (if it exists) is undetermined as $\Gamma + Z$ satisfies \eqref{Eq:Hamiltonian equation} for any $Z \in K$.
A sufficient condition for the existence of a Hamiltonian vector field $X_H \,=\, \Gamma$ is clearly  that $Z(H) \,=\, 0$, for all $Z \in K$. 
When such condition is fulfilled, we say that the dynamics is global on $\mathcal M$. 
When this is not the case, the relation 
$$
Z(H)=0 \, ,
$$
for any $Z\,\in\,K$ selects a subset of $\mathcal M$ on which a vector field $\Gamma$ can be obtained. 
The further condition that $\Gamma$ is tangent to such subset may provide a further constraint. 
This is the first step of the so calles pre-symplectic constraint algorithm (PCA) \cite{Gotay-Nester-Hinds1978-DiracBergmann_constraints}.
The subset (which is assumed to be a submanifold) $\mathcal{M}_1 \,=\, \left\{\, m \in \mathcal{M} \;\; : \;\; Z_m(H) \,=\,0 \;\; \forall \,\, Z \in K \,\right\}$ is called the primary constraint submanifold of the pre-symplectic Hamiltonian system $(\mathcal{M},\, \omega,\, H)$.
The restriction of $\omega$ and $H$ to $\mathcal{M}_1$ determines a new pre-symplectic system provided that $\mathcal{M}_1$ is a smooth submanifold of $\mathcal{M}$, that we denote by  $(\mathcal{M}_1,\, \omega_1,\, H_1)$.
Denoting by $K_1$ the characteristic distribution of $\omega_1$ we define recursively
\be\label{eq:pca}
\begin{split}
\mathcal{M}_{k+1} \,&=\, \left\{\, m \in \mathcal{M}_k \;\; : \;\; Z_m(H_k)\,=\, 0 \;\; \forall \,\, Z \in K_k \,\right\} \,\,\, k \,=\, 1, 2, ... \\
\omega_{k+1} \,&=\, \omega_k\bigr|_{\mathcal{M}_{k+1}} \,; \\  
H_{k+1} \,&=\, H_k\bigr|_{\mathcal{M}_{k+1}} \,;  \\ 
K_{k+1} \,&=\, \mathrm{ker}\omega_{k+1}
\end{split}
\ee
At each step of the previous process it will be assumed that $\mathcal M_k$ is a manifold, so that $(\mathcal{M}_k,\, \omega_k)$ is a (pre-symplectic) manifold. 
Often it is assumed that the PCA stops after a finite number of steps, i.e., $\mathcal{M}_r \,=\, \mathcal{M}_{r+1}$ for some $r$, implying that $\omega_r \,=\, \omega_{r+1}$.
The manifold $(\mathcal{M}_r,\, \omega_r)$ will be called the final constraints manifold.
It follows that the system $(\mathcal{M}_r,\, \omega_r,\, H_r)$ is consistent, that is, there is a solution to \eqref{Eq:Hamiltonian equation} all over $\mathcal{M}_r$.
In the subsequent analysis it will be assumed that our pre-symplectic system is a consistent Hamiltonian system obtained after a PCA analysis if needed.

\paragraph{Action of Lie groups on pre-symplectic manifolds and the momentum map.} 
Consider a Lie group acting on $\mathcal{M}$, namely  consider a Lie group $\mathcal{G}$ together with a representation of it on $\mathcal{M}$
\be
\Phi \;\; : \;\; \mathcal{G} \to \mathcal{T}(\mathcal{M}) \;\; : \;\;  g \mapsto \Phi_g
\ee
where $\mathcal{T}(\mathcal{M})$ is the set of maps from $\mathcal{M}$ to itself.
Such a map is a representation if it is a homomorphism, i.e.  if
\be
\Phi_{g \cdot h} \,=\, \Phi_g \circ \Phi_h
\ee
where $\cdot$ is the composition law of $\mathcal{G}$ and $\circ$ is the composition of two maps in $\mathcal{T}(\mathcal{M})$.
We will refer to $\Phi_g$ as the action of $\mathcal{G}$ on $\mathcal{M}$.
It will be assumed that the action is smooth, that is that the map $\Phi_m \colon \mathcal{G} \to \mathcal{M}$, given by $\Phi_m(g) = \Phi_g(m)$ is smooth for all $m\in \mathcal{M}$.
The action of a Lie group on $(\mathcal{M}, \, \omega)$ is said to be canonical if $\Phi^\star_g \omega \,=\, \omega \;\;\; \forall \,\, g\in \mathcal{G}$.
Equivalently, the action is canonical if $\mathfrak{L}_{X_\xi} \omega \,=\, 0 \;\;\; \forall \,\, \xi \in \mathfrak{g}$ where $X_\xi$ is the Killing vector field on $\mathcal{M}$ corresponding to the element $\xi$ in the Lie algebra $\mathfrak{g}$ of $\mathcal{G}$.
The action is said to be Hamiltonian if a function $J_\xi$ on $\mathcal{M}$ exists such that
\be
i_{X_\xi} \omega \,=\, \dd J_\xi \,.
\ee
It is easy to see  that if the action of $\mathcal{G}$ is Hamiltonian then it is canonical. The converse is not true: there exist canonical actions which are not Hamiltonians (the form $i_{X_\xi} \omega$ is closed, not necessarily exact).  
In this case the action is also said to be weakly Hamiltonian.

When the action of $\mathcal{G}$ on $\mathcal{M}$ is Hamiltonian, the map $\mathbb{J}$ defined by
\be
\mathbb{J} \;\; : \;\; \mathcal{M} \to \mathfrak{g}^\star \;\; :\;\; \langle \mathbb{J}(m),\, \xi \rangle \,=\, J_\xi(m) \;\; m \in \mathcal{M},\, \xi \in \mathfrak{g}
\ee
is well defined and is called the momentum map of the action of $\mathcal{G}$ on $(\mathcal{M},\, \omega)$.
Here, $\mathfrak{g}^\star$ denotes the dual of the Lie algebra of $\mathcal{G}$.
If the Lie algebra is infinite-dimensional, $\mathfrak{g}^\star$ is the topological dual.

The functions $J_\xi$ can be determined up to constants. 
Even more, the identity $[X_\xi,\, X_\zeta] \,=\, X_{[\xi,\,\zeta]}$ for $\xi,\, \zeta \in \mathfrak{g}$ implies that the map $\mathbb{J}$ is equivariant  up to a $\mathfrak{g}^\star$-valued $1$-cocycle on $\mathcal{G}$ with respect to the co-adjoint action of $\mathcal{G}$ on $\mathfrak{g}^\star$ (see \cite[Prop. 4.5.21]{Ortega-Ratiu2004-Momentum_map}).
If the momentum map $\mathbb{J}$ is equivariant, i.e. if  
\be \label{Eq:equivariance action G}
\mathbb{J}(\Phi_g \cdot m) \,=\, \mathrm{Ad}_g^\star (\mathbb{J}(m)) \;\; \qquad \forall g \in \mathcal{G},\, m \in \mathcal{M} \,,
\ee
we  say that the action of $\mathcal{G}$ on $(\mathcal{M},\, \omega)$ is strongly Hamiltonian.

The Lie group $\mathcal{G}$ is said to be a symmetry group for the Hamiltonian system $(\mathcal{M},\,\omega,\, H)$ if the action of $\mathcal{G}$ on $(\mathcal{M},\, \omega)$ is canonical (or, a fortiori, Hamiltonian/strongly Hamiltonian) and if $\mathcal{G}$ is a symmetry for $H$, i.e., $H_{\Phi_g ( m)}\,=\, H_m$, for all $g \in \mathcal{G},\,\,\, \forall \,\, m \in \mathcal{M}$. 
The latter condition is equivalent to say that $H$ is invariant with respect to the action of $\mathcal{G}$ on $\mathcal{M}$, i.e., $X_\xi (H) \,=\, 0 \;\;\; \forall \,\, \xi \in \mathfrak{g}$.
The converse also holds, i.e., $X_\xi(H) \,=\,0 \;\;\; \forall\,\, \xi \in \mathfrak{g}$ implies that $\mathcal{G}$ is a symmetry for $H$, provided that $\mathcal{G}$ is connected.

If $\mathcal{G}$ acts canonically on the pre-symplectic manifold $(\mathcal{M},\, \omega)$, it preserves the characteristic foliation, i.e.
\be \label{Eq:conservation of characteristic foliation}
{\Phi_g}_\star K_m \,\subseteq\, K_{\Phi_g( m)} \;\;\; \forall \,\, g \in \mathcal{G} \,, \;\; \forall \,\, m \in \mathcal{M} \,.
\ee
Then, the $\mathcal{G}$-invariance of $K$ implies that $[X_\xi,\, K] \subseteq K$ $\forall \,\, \xi \in \mathfrak{g}$.

Now, let $F$ and $G$ be two functions on $\mathcal{M}$ such that there exist vector fields $X_F$ and $X_{G}$ such that $i_{X_F} \omega \,=\, \dd F$ and $i_{X_{G}} \omega \,=\, \dd G$ (i.e., such that $Z(F) \,=\, Z(G) \,=\, 0$ $\forall \,\, Z \in K$). 
We define the (pre-symplectic) Poisson bracket between $F$ and $G$ as
\be
\{F,\, G\} \,=\, \omega(X_F,\, X_{F'}) \,=\, X_G(F) \,=\, -X_F(G) \,.
\ee
It turns out that this bracket is well-defined, bilinear, satisfies the Jacobi identity (because the map $\xi \mapsto X_\xi$ is a homomorphism of Lie algebras) and the  Leibnitz' rule (because $\dd(FF') \,=\, (\dd F) F' + F \dd F'$).
Moreover, if the action of $\mathcal{G}$ on $(\mathcal{M},\,\omega)$ is strongly Hamiltonian, then it follows from \eqref{Eq:equivariance action G} that 
\be
\{J_\xi,\, J_\zeta\} \,=\, J_{[\xi,\,\zeta]}
\ee
i.e., the components of the momentum map close  a Lie algebra, called the algebra of currents of the symmetry.

If $\mathcal{G}$ is a strongly Hamiltonian symmetry group of the pre-symplectic Hamiltonian system $(\mathcal{M},\, \omega,\, H)$, we get
\be
0 \,=\, X_\xi(H) \,=\, i_{X_\xi} \dd H \,=\, \omega(X_H,\, X_\xi) \,=\, \{H,\, J_\xi\} \,=\, -\{J_\xi,\, H\}\,=\, -i_{X_H}i_{X_\xi} \omega \,=\, -X_H(J_\xi) \,=\, 0 
\ee
i.e.,
\be
X_\xi (H) \,=\, -X_H(J_\xi) \,=\, 0
\ee
which is the particular form of Noether's theorem in this setting (any strongly Hamiltonian infinitesimal symmetry, $X_\xi(H)\,=\,0$, defines a constant of the motion, $X_H(J_\xi) \,=\,0$, and vice-versa).

The sections $Z$ of the characteristic distribution $K$ define a Lie algebra (possibly infinite-dimensional).
If the Lie algebra $\Gamma(K)$ can be integrated, i.e., there is a group $\mathcal{K}$ (possibly infinite-dimensional) whose Lie algebra is $\Gamma(K)$, then it will be called the gauge group of transformations of the theory (that in some cases can be identified with the group of, geometric, gauge transformations of the theory, see Sect. \ref{Sec:Symmetry theory on the Euler-Lagrange space}).

Functions which are constant along the leaves of the characteristic distribution are called the (classical) observables of the theory and, if the quotient space $\mathcal{M}/K$ is a manifold, they are in one-to-one correspondence with smooth functions on $\mathcal{M}/K$, i.e., $K(F) \,=\,0$ iff $F \,=\, \mathrm{pr} f$ for some $f \;\; : \;\; \mathcal{M}/K \to \mathbb{R}$.
Moreover, in such situation $$\{F,\,G\} \,=\, \mathrm{pr}^\star \{f,\,g\},$$ where $F \,=\, \mathrm{pr}^\star f$ and $G\,=\,\mathrm{pr}^\star g$, while  $\{f,\,g\}$ is the Poisson bracket on $\mathcal{M}/K$ with respect to the induced symplectic structure on $\mathcal{M}/K$.

Getting  rid of the gauge degrees of freedom amounts to pass to the quotient space $\mathcal{M}/K$ (note that the canonical action of the group $\mathcal{G}$ passes to the quotient too because of \eqref{Eq:conservation of characteristic foliation}, and the induced action on the quotient symplectic manifold is canonical/Hamiltonian/strongly Hamiltonian respectively).
Since such a quotient is in general quite singular (one of the most interesting way to deal with such singularity is to study it within a cohomological approach \cite{Asorey-Mitter1986-Cohomology}), we prefer here to consider a suitable symplectic realization  pre-symplectic manifold $(\mathcal{M},\, \omega)$.
This is the content of the following section. 

\paragraph{The coisotropic embedding theorem.}
To this aim, we analyse  an equivariant version of the coisotropic embedding theorem stated by \textit{M. Gotay} \cite{Gotay-Sniatycki1981-Coisotropic_embeddings, Guillemin-Sternberg1990-Symplectic_techniques}.
The main idea is that, given a pre-symplectic manifold $(\mathcal{M},\omega)$, it is possible to construct, in a canonical way, a symplectic manifold $(\mathcal{P},\, \omega_{\mathcal{P}})$ and an embedding $\mathfrak{i} \;\; : \;\; \mathcal{M} \hookrightarrow \mathcal{P}$, such that $(\mathcal{M},\, \omega)$ is a coisotropic submanifold of $(\mathcal{P},\, \omega_{\mathcal{P}})$, i.e., $\mathfrak{i}^\star \omega_{\mathcal{P}}\,=\, \omega$.
Such a manifold $\mathcal{P}$ can be constructed explicitly as a tubular neighborhood of the zero section of the dual of the characteristic bundle of $\mathcal{M}$, $\mathbf{K}$, i.e. of the bundle $\mathbf{K}^\star \to \mathcal{M}$.
Note that the tangent bundle of $\mathbf{K}^\star \to \mathcal{M}$ along the zero section, that we denote by $\mathbf{T}_{\mathcal{M}} \mathbf{K}^\star$, can be decomposed as
\be 
\mathbf{T}_{\mathcal{M}}\mathbf{K}^\star \,=\, \mathbf{K}^\star \oplus \mathbf{T}\mathcal{M}
\ee
i.e., at each point $(m,\,0) \,=\, 0_m$, we have 
\be
\mathbf{T}_{0_m}\mathbf{K}^\star \simeq K_m^\star \oplus \mathbf{T}_m \mathcal{M} \,=\, K_m^\star \oplus K_m \oplus W_m
\ee
where $W_m$ is a complementary subspace to $K_m \subset \mathbf{T}_m \mathcal{M}$, i.e., $\mathbf{T}_m \mathcal{M} \,=\, K_m \oplus W_m$.
Then $W_m$ is a symplectic subspace of $(\mathbf{T}_{m}\mathcal{M},\, \omega_m)$ and we can define a (linear) symplectic form on $\mathbf{T}_{0_m}\mathbf{K}^\star \,=:\, \mathbf{T}_{0_m}\mathcal{P}$ as follows 
\be
\begin{split}
\omega_{\mathcal{P}}(W \oplus Z \oplus \mu,\, W' \oplus Z' \oplus \mu') &\,=\, \omega(W,\, W') + \langle \mu,\, Z'\rangle - \langle \mu',\, Z \rangle \\  & W,\,W' \in W_m,\, Z,\,Z' \in K_m,\, \mu,\, \mu' \in K^\star_m \,.
\end{split}
\ee
Such linear symplectic form along the subbundle $\mathbf{T}_{\mathcal{M}}\mathbf{K}^\star$ can be extended to a symplectic form on a tubular neighborhood of the zero section of $\mathbf{K}^\star \to \mathcal{M}$ and it gives the symplectic form we are looking for (note that $\mathcal{M} \hookrightarrow \mathcal{P} \subset \mathbf{K}^\star$ is canonically embedded in $\mathbf{K}^\star$, and $\omega_{\mathcal{P}}\bigr|_{\mathcal{M}} \,=\, \omega$).
Local coordinates on $(\mathcal{P},\, \omega_{\mathcal{P}})$ can be chosen as follows.
Let $\{E_a\}$ be a local frame on $K$, then $\mu \in K^\star$ can be written as $\mu \,=\, \mu_a E^a$, where $E^a$ is the dual frame of $E_a$.
Around any point on a regular leaf of the foliation $K$ we can find coordinates $(w^\alpha,\, Z^a)$ such that $w^\alpha$ are transverse coordinates and $Z^a$ are such that $Z\,=\, Z^a E_a$ is a tangent vector in $K$.
Then we get 
\be 
\omega_{\mathcal{P}} \,=\, \omega_{\alpha \beta} \dd w^\alpha \wedge \dd w^\beta + \dd Z^a \wedge \dd \mu_a \,.
\ee
Moreover, the embedding $\mathfrak{i} \;\; : \;\; \mathcal{M} \hookrightarrow \mathcal{P}$ is described by $\mu_a \,=\, 0$.

If the action of $\mathcal{G}$ on $(\mathcal{M},\, \omega)$ is strongly Hamiltonian, then there is a lift of such action to $\mathbf{K}^\star$.
The lift is explicitly described as follows.
For each $g \in \mathcal{G}$, consider the map $Tg_m \;\; : \;\; K_m \to K_{\Phi_g \cdot m}$ given by 
$$
Tg_m(Z) \,=\, \frac{d}{ds}\biggr|_{s=0} \Phi_g \cdot \gamma(s) \, ,
$$ 
with $\gamma$ a curve in $K$ and $\dot{\gamma}(0)\,=\, Z_m$, i.e., as the distribution $K$ is $\mathcal{G}$-invariant, the tangent map ${\Phi_g}_\star \,=\, Tg$ maps $K$ into itself.
The dual map $Tg^\star_m \;\; : \;\; K_{\Phi_g \cdot m}^\star \to K_m^\star$, is defined as
\be
\langle Tg_m^\star \mu_{\Phi_g\cdot m},\,Z_m \rangle \,=\, \langle \mu_{\Phi_g \cdot m},\, Tg_m Z_m \rangle \;\;\; \forall \,\, Z_m \in K_m \,.
\ee
Then we define the lift of the action of $\mathcal{G}$ to $\mathbf{K}^\star$ as
\be
\tilde{\Phi}_g (m,\,\mu) \,=\, (\Phi_g (m) ,\, (Tg^{-1})^\star \mu) \;\;\; g \in \mathcal{G},\;\; (m,\,\mu)\in \mathbf{K}^\star \,.
\ee
If the tubular neighborhood $\mathcal{P}$ can be chosen to be $\mathcal{G}$-invariant (for instance, if there is a $\mathcal{G}$-invariant metric on $\mathcal{M}$), then the induced action of $\mathcal{G}$ on $(\mathcal{P},\, \omega_{\mathcal{P}})$ is strongly Hamiltonian if the action of $\mathcal{G}$ is.
The best way to see that is by explicitly constructing an equivariant momentum map for such action.
We do it upon  introducing the technical requirement that the action of $\mathcal{G}$ on $\mathcal{M}$ is  quasi-free, i.e.  the isotropy group $\mathcal{G}_m$ is finite (or discrete).
If this is the case, then the map $\mathfrak{i}_m \;\; : \;\; \mathfrak{g} \to \mathbf{T}_m \mathcal{M}$, that maps the element $\xi \in \mathfrak{g}$ into the tangent vector $X_\xi(m) \in \mathbf{T}_m \mathcal{M}$ is injective and we can identify $\mathfrak{g}$ with its image $\mathfrak{i}_m(\mathfrak{g}) \subset \mathbf{T}_m \mathcal{M}$.
Denoting by $\mathfrak{g}_m \,:=\, \mathfrak{i}_m \mathfrak{g}$, we can construct the subbundle $\tilde{\mathfrak{g}} \,=\, \bigcup_{m \in \mathcal{M}} \mathfrak{g}_m \subset \mathbf{T}\mathcal{M}$, where fibres are copies of the Lie algebra $\mathfrak{g}$.
Then, we can consider the intersection $K \bigcap \tilde{\mathfrak{g}}$, i.e., at each point $m\in\mathcal{M}$, $\left( K \bigcap \tilde{\mathfrak{g}} \right)_m \,=\, K_m \bigcap \mathfrak{g}_m$.
Assuming that $K \bigcap \tilde{\mathfrak{g}} \subset \mathbf{T}\mathcal{M}$ is again a subbundle, we can denote it by $\tilde{\mathfrak{g}}_{\textsc{gauge}}\,=\,K \bigcap \tilde{\mathfrak{g}}$.
Then we have the short exact sequence of bundles
\be
\begin{tikzcd}
0 \to \tilde{\mathfrak{g}}_{\textsc{gauge}} \to \tilde{\mathfrak{g}} \to \tilde{\mathfrak{g}} / \tilde{\mathfrak{g}}_{\textsc{gauge}} \to 0 \,.
\end{tikzcd}
\ee
The space of sections of $\tilde{\mathfrak{g}} \to \mathcal{M}$ carries a Lie algebra structure as $[\tilde{\xi},\, \tilde{\zeta}]_m \,=\, [\tilde{\xi}_m,\, \tilde{\zeta}_m]$, where $\tilde{\xi}_m$ and $\tilde{\zeta}_m \in \mathfrak{g}_m$, are identified with the corresponding elements in $\mathfrak{g}$ (by means of $\mathfrak{i}_m$) and the bracket is the Lie algebra bracket in $\mathfrak{g}$.
The space of sections of $\tilde{\mathfrak{g}}_{\textsc{gauge}}$ is a Lie algebra ideal of $\tilde{\mathfrak{g}}$ and the quotient $\tilde{\mathfrak{g}}/\tilde{\mathfrak{g}}_{\textsc{gauge}}$ becomes a Lie algebra again.
We may denote the underlying bundle $\tilde{\mathfrak{g}}_{\textsc{dyn}}$, and, at each point $m \in \mathcal{M}$, it defines a Lie subalgebra of $\mathfrak{g}$ (that could vary with $m$).
A $(\mathcal{G},K)$-connection on $(\mathcal{M},\,\omega)$ will be an equivariant bundle map $A \;\; : \;\; \tilde{\mathfrak{g}} \to \tilde{\mathfrak{g}}_{\textsc{gauge}}$, i.e. $A_m \;\; : \;\; \tilde{\mathfrak{g}}_m \to \tilde{\mathfrak{g}}_{\textsc{gauge}, m}$ and $$(\mathrm{Ad}_g)_\star (A_m(\tilde{\xi}_m)) \,=\, A_{\Phi_g \cdot m}((\mathrm{Ad}_g)_\star \tilde{\xi}_m),$$  for any $ g \in \mathcal{G}$, $m \in \mathcal{M}$, $\tilde{\xi}_m \in \mathfrak{g}_m$\footnote{Note that such a map always exists, it can be constructed using local trivializations and summing them using a partition of unity.}.
Then, we construct the map $\mathbb{J}_\mathcal{P} \;\; : \;\; \mathcal{P} \to \mathfrak{g}^\star$ as
\be \label{Eq: momentum map coisotropic embedding}
\langle \mathbb{J}_\mathcal{P}(m,\,\mu), \xi \rangle \,=\, \langle \mathbb{J}(m),\, \xi \rangle + \langle \mu,\, A_m(\alpha_\xi) \rangle  \;\;\; \forall \,\, m \in \mathcal{M},\,\, \mu \in K^\star,\,\, \xi \in \mathfrak{g}\,.
\ee
One can now show that  $\mathbb{J}_{\mathcal{P}}$ is equivariant, i.e.
\be
\mathbb{J}_\mathcal{P}(\tilde{\Phi}_g \cdot (m,\,\mu))\,=\, \mathrm{Ad}_{g^{-1}}^\star(\mathbb{J}_\mathcal{P}(m,\,\mu))
\ee
and the action of $\mathcal{G}$ on $\mathcal{P}$ is strongly Hamiltonian.


\section{The Euler-Lagrange space as pre-symplectic manifold}
\label{Sec:The Euler-Lagrange space as pre-symplectic manifold}

It is well known that the space of solutions of a first order Hamiltonian field theory can be canonically equipped with a pre-symplectic structure \cite{Garcia-Perez-Rendon1969-Symplectic_Quantized_FieldsI, Garcia-Perez-Rendon1969-Symplectic_Quantized_FieldsII, Marsd-Mont-Morr-Thomp1986-Covariant_Poisson_bracket, Crnkovic-Witten1986-Covariant_canonical_formalism, Crnkovic1988-Symplectic_Cov_Phase_Space, Forger-Romero2005-Covariant_poisson_brackets, Forger-Salles2015-Covariant_Poisson_brackets, Montiel-Molgado-Lopez2021-Geometric_field_theory_Gravity, Margalef-Villasenor2021-Covariant_phase_space_boundary, Gieres2021-Covariant_field_theories}.
In this section we briefly explain the main steps of the construction in order to introduce the setting to which we will apply the abstract theory developed in Sect. \ref{Sec:Symmetries for pre-symplectic dynamics} to fix the notations and the relevant geometrical structures so that this contribution will be as self-consistent as possible. 
We refer to \cite{Ibort-Spivak2017-Covariant_Hamiltonian_YangMills, Asorey-Ibort-Spivak2017-Boundary_conditions, Ciaglia-DC-Ibort-Marmo-Schiav2020-Jacobi_Particles, Ciaglia-DC-Ibort-Marmo-Schiav2020-Jacobi_Fields} and to a companion paper \cite{Ciaglia-DC-Ibort-Marmo-Schiav-Zamp2021-Cov_brackets_toappear} for all the technical details and for many proofs.
\vsp

In the multisymplectic formalism, a first order Hamiltonian field theory is specified by:
\begin{itemize}
\item A fibre bundle over an underlying (orientable) space-time $\mathscr{M}$ of dimension $1 +d$ with smooth boundary $\partial\mathscr{M}$, say $\pi \;\; : \;\; \mathbb{E} \to \mathscr{M}$, whose sections $\phi$ represent the configuration fields of the theory.
The carrier space where the description of the field theory takes place is the so called Covariant Phase Space\footnote{It is worth pointing out that this terminology is not standard and that many authors in the literature refers to Covariant Phase Space as the space of solutions of the equations of the motion.
Instead, we will refer to it as the \textsc{Euler-Lagrange space}.}.
It is defined as the reduced dual\footnote{In the standard terminology of affine bundles \cite{Rossi-Saunders2014-Dual_Jet_Bundles, Ibort-Spivak2017-Covariant_Hamiltonian_YangMills}} of the first order jet bundle of the fibration $\pi$ and it is denoted by $\mathcal{P}(\mathbb{E})$.
Here, we only recall that given a chart on $\mathscr{M}$, $(U_{\mathscr{M}}, \psi_{U_{\mathscr{M}}})$, $\psi_{U_{\mathscr{M}}}(m) \,=\, (x^0,...,x^d)\,=:\, x$ (with $m \in U_{\mathscr{M}}\subset \mathscr{M}$) and a fibered chart on $\pi$, $(U_{\mathbb{E}}, \psi_{U_{\mathbb{E}}})$, $\psi_{U_{\mathbb{E}}}(\mathfrak{e})\,=\, (x^0,...,x^d, u^1,...,u^r)\,=:\, (x, u)$ (with $\mathfrak{e} \in U_{\mathbb{E}}\subset \mathbb{E}$ and $r$ being the dimension of the fiber of $\pi$), an adapted fibered chart can be given on the first order jet bundle of $\pi$, $\mathbf{J}^1 \pi$, say $(U_{\mathbf{J}^1 \pi}, \psi_{U_{\mathbf{J}^1 \pi}})$, $\psi_{U_{\mathbf{J}^1 \pi}}(\mathfrak{e}_x) \,=\, (x^0,...,x^d, u^1,..., u^r, z^1_0,..., z^r_d)\,=:\, (x, u, z)$ (with $\mathfrak{e}_x \in U_{\mathbf{J}^1 \pi} \subset \mathbf{J}^1 \pi$).
As well as $\mathbf{J}^1 \pi$, also $\pe$ is a fibre bundle both over $\mathbb{E}$ and over $\mathscr{M}$ where an adapted fibered chart can be given as $(U_{\mathcal{P}(\mathbb{E})}, \psi_{U_{\mathcal{P}(\mathbb{E})}})$, $\psi_{U_{\mathcal{P}(\mathbb{E})}}(\mathfrak{p}) \,=\, (x^0,...,x^d, u^1,...,u^r, \rho^0_1,..., \rho^d_r) \,=:\, (x, u, \rho)$ (with $\mathfrak{p} \in U_{\mathcal{P}(\mathbb{E})} \subset \mathcal{P}(\mathbb{E})$) where the coordinates $\rho^\mu_a$ represent the ``dual coordinates'' of the fibered coordinates $z^a_\mu$ on $\mathbf{J}^1\pi$.
\be
\begin{tikzcd}
\mathbf{J}^1 \pi \arrow[ddr, "\pi_1", bend right = 20, swap] \arrow[dr, "\pi^1_0"] & & \mathcal{P}(\mathbb{E}) \arrow[ddl, "\delta_1", bend left = 20] \arrow[dl, "\delta^1_0", swap] \\
 & \mathbb{E} \arrow[d, "\pi"] & \\
 & \mathscr{M} \arrow[u, "\phi", dashed, bend left = 30] &
\end{tikzcd}
\ee
\item A \textsc{Hamiltonian} function that, for all the purposes of the present paper, can be considered as a local function on $\mathcal{P}(\mathbb{E})$
\be
H \;\; : \;\; U_{\mathcal{P}(\mathbb{E})} \to \mathbb{R} \;\; :\;\; \psi^{-1}_{U_{\mathcal{P}(\mathbb{E})}}(x, u, \rho) \mapsto H(\psi^{-1}_{U_{\mathcal{P}(\mathbb{E})}}(x, u, \rho)) \,.
\ee
Throughout the paper, with the usual  (slight) abuse of notation, we will use $H$ to denote $H \circ \psi^{-1}_{U_{\mathcal{P}(\mathbb{E})}}$.
\end{itemize}
The  space of fields used to describe the dynamical content of the theory within the Hamiltonian formalism is the space of sections of $\delta_1$ which factorizes into a section of $\pi$, say $\phi$, and a section of $\delta^1_0$, say $P$.
We denote elements of this space as ordered pairs $(\phi, P)$.
Note that
\begin{equation}\label{eq:factorization}
\chi \,=\, P \circ \phi \, ,
\end{equation}
is a section of $\delta_1$.  
Often we will denote the pair $(\phi, P)$ by $\chi$ itself.
On this space of sections we will define the action functional encoding the dynamical content of the theory.
For technical reasons, it would be worth working on a space of fields admitting at least a Banach manifold structure but this is not the case for the space of smooth sections of a fibre bundle which is a Fr\'echet space.
For this reason, in all the examples considered we will complete the space of fields with respect to a suitable Banach norm (actually, in all the cases considered it will be a Sobolev norm) in which the action functional is a continuous one.
In this way, the action functional can be extended by continuity to such a completion which turns to be our new space of fields. 
We  call this space the space of dynamical fields and we denote it by $\mathcal{F}_{\mathcal{P}(\mathbb{E})}$\footnote{The corresponding subset of $\Gamma^\infty(\pi)$ of configuration fields is denoted by $\mathcal{F}_\mathbb{E}$.}.
In developing the abstract theory we will consider the action functional to be well defined on the Banach manifold $\fpe$ and we will take care of constructing the proper completion case by case in the examples.

The Covariant Phase Space admits a canonical $(d+1)$-form whenever a Hamiltonian is fixed which, in local adapted coordinates, takes the form:
\be
\Theta_H \,=\, \rho^\mu_a \dd u^a \wedge i_\mu \, vol_{\mathscr{M}} - H vol_{\mathscr{M}} \, ,
\ee
where $vol_{\mathscr{M}}$ is a volume form on $\mathscr{M}$ and $i_\mu$ denotes the contraction with the vector field $\frac{\partial}{\partial x^\mu}$.
In terms of such a differential form, an action functional on the space of dynamical fields can be defined in the following way:
\be
\mathscr{S}_\chi \,=\, \int_{\mathscr{M}} \chi^\star \, \Theta_H \,.
\ee
The first variation of $\mathscr{S}$ along the direction $\mathbb{X}_\chi \in \mathbf{T}_\chi \mathcal{F}_{\mathcal{P}(\mathbb{E})}$ reads
\be \label{Eq:first variational formula}
\delta_{\mathbb{X}_\chi} \mathscr{S}_\chi \,=\, \int_{\mathscr{M}} \chi^\star \left[\, i_X \dd \Theta_H \,\right] + \int_{\partial \mathscr{M}} \chi_{\partial \mathscr{M}}^\star \left[\, i_X \Theta_H \,\right] \,,
\ee
where $\chi_{\partial \mathscr{M}} \,=\, \chi\bigr|_{\partial \mathscr{M}} \,=\, \mathfrak{i}_{\partial \mathscr{M}}^\star \chi \in \fpe\bigr|_{\de \m} \,=:\, \fpe^{\de \m}$, $\mathfrak{i}_{\partial \mathscr{M}}$ denoting the canonical immersion of $\partial \mathscr{M}$ into $\mathscr{M}$, and $X$ is a vector field over an open neighborhood of the image of $\chi$ which restricts to $\mathbb{X}_\chi$ when evaluated along $\chi$.
Equation \eqref{Eq:first variational formula} is known as first variational formula.

The first variational formula can be given an interpretation in terms of differential forms on $\fpe$ in the following way.
The term on the left hand side is the differential of the function (on $\fpe$) $\ac$ contracted along the tangent vector $\mathbb{X}_\chi \in \tfpe$, i.e. what we can write as $i_{\mathbb{X}_\chi} \dd \ac$.
Regarding the right hand side, also the first term can be interpreted as the contraction of a differential $1$-form on $\fpe$ (seen as an application from each $\tfpe$ to $\mathbb{R}$) with the tangent vector $\mathbb{X}_\chi$.
We  call such a differential $1$-form the Euler-Lagrange form and we  denote it by $\el$, 
\be
i_{\mathbb{X}_\chi}\el_\chi \,=\, \int_{\m} \chi^\star \left[\, i_X \dd \Theta_H \,\right] \,.
\ee
Regarding the last term on the right hand side, again it can be seen as the contraction of a differential form with a tangent vector.
However, since it only depends on the restriction of $\chi$ to the boundary $\partial \m$, if we want to interpret it in terms of a differential form on $\fpe$, it must be a form which is the pull-back of a differential form on $\fpe^{\de \m}$ via the restriction map
\be\label{eq:partialM}
\Pi_{\de \m} \;\; : \;\; \fpe \to \fpe^{\de \m} \;\; : \;\; \chi \mapsto \chi\bigr|_{\de \m} \,.
\ee
We denote such a differential form by $\Pi_{\de \m}^\star \alpha^{\de \m}$
\be
i_{\mathbb{X}_\chi} \Pi^\star_{\de \m} \alpha^{\de \m} \, =\, \int_{\de \m} \chi_{\de \m}^\star \left[\, i_X \dd \Theta_H \,\right] \,.
\ee
A few more comments regarding the latter differential form are worth to be made.
Indeed, it is exactly the differential form which will make the space of the solutions of the theory into a pre-symplectic manifold.
First, let us note that by virtue of the embedding theorem, we can always choose the coordinate system on $\m$ around $\de \m$ such that $x^0$ is the coordinate transversal to $\de \m$.
Consequently, the space of fields restricted to $\de \m$, say $\fpe^{\de \m}$, splits as follows:
\be
\fpe^{\de \m} \,=\, \left\{\, \phi^a \bigr|_{\de \m},\, P^\mu_a \bigr|_{\de \m} \, \right\} \,=\, \left\{\,\phi^a \bigr|_{\de \m},\, P^0_a \bigr|_{\de \m},\, P^j_a \bigr|_{\de \m} \,\right\}_{j=1,...,d} \,=:\, \left\{\, \varphi^a,\, p_a,\, \beta^j_a  \,\right\} \,=\, \mathcal{F}_{\pe, 0}^{\de \m} \times \mathscr{B}^{\de \m} \,.
\ee
Now, the space $\mathcal{F}_{\pe, 0}^{\de \m}$ is isomorphic with the cotangent bundle of the space of sections of the bundle $\mathbb{E}$ restricted to $\partial \mathscr{M}$, $\mathcal{F}_{\mathbb{E}}\mid_{\partial \mathscr{M}}$, denoted as $\mathcal{F}_{\mathbb{\partial E}}$ and, denoting by $\tau$ the projection from $\fpe^{\de \m}$ to $\mathbf{T}^\star \mathcal{F}_{\mathbb{E}}$, it is a direct computation to show that:
\be
-\dd \Pi_{\de \m}^\star \alpha^{\de \m} \,=:\, \Pi_{\de \m}^\star \Omega^{\de \m} \, =\, \tau^\star \omega \, ,
\ee
where $\omega = - \dd \alpha^{\de \m}$, is the canonical symplectic structure of $\mathbf{T}^\star  \mathcal{F}_{\partial \mathbb{E}}$.
Explicitly it reads
\be 
\Pi_{\de \m}^\star \Omega^{\de \m}(\mathbb{X},\, \mathbb{Y}) \,=\, \int_{\de \m} \left[\, \mathbb{X}_{\varphi}^a {\mathbb{Y}_p}_a - {\mathbb{X}_p}_a \mathbb{Y}_\varphi^a \,\right] vol_{\de \m}
\ee 
which, in the particular instance of particle dynamics, becomes the standard differential form $\dd q^j \wedge \dd p_j$ of the Hamiltonian formalism \cite{Ciaglia-DC-Ibort-Marmo-Schiav2020-Jacobi_Particles}.

In terms of the differential forms on $\fpe$ defined above, the first variational formula  reads
\be
\dd \ac_\chi \,=\, \el_\chi + \Pi^\star_{\de \m} \alpha^{\de \m}_\chi \,.
\ee
The dynamical content of the theory, i.e. the equations of the motion,  can be obtained via a variational principle \'a la Schwinger-Weiss \cite{Schwinger1970-Quantum_kinematics_dynamics, Ibort-Marmo-Asorey-Falceto}.
It states that the solutions  of the theory considered are those $\chi$ for which the first variation of the action functional only depends on boundary terms, i.e., on terms that only depends on the restrictions of the fields to the boundary $\de \m$.
Therefore, by looking at the first variational formula, the solutions of the theory are those for which
\be
\el_\chi \,=\, 0 \,.
\ee
In this perspective the solutions of the equations of the motion are the zeroes of a differential form (the Euler-Lagrange form) on an infinite-dimensional manifold.
From now on, we look at  the space of solutions of the equations of the motion as the space of zeroes of $\el$ and we denote it by $\elag_{\m}$:
\be
\begin{split}
\elag_{\m} \,&=\, \left\{\, \chi \in \fpe \;\; : \;\; \el_\chi \,=\,0 \,\right\} \,=\, \left\{\, \chi \in \fpe \;\; : \;\; \chi^\star \left[\, i_X \dd \Theta_H \,\right] \,=\, 0 \;\;\; \forall \,\, X \in \mathfrak{X}^v(U^{(\chi)}) \,\right\} \,=\, \\ 
\,&=\, \left\{ \chi \in \fpe \;\; : \;\; \frac{\de \phi^a}{\de x^\mu} \,=\, \frac{\de H}{\rho^\mu_a}\biggr|_{\chi} \,\,\, \text{and} \,\,\, \frac{\de P^\mu_a}{\de x^\mu} \,=\, -\frac{\de H}{\de u^a}\biggr|_\chi \, \right\} \,.
\end{split}
\ee

Even if it nothing is said about the differential structure of $\elag_\m$ we will always assume from now on $\elag_\m$ to be a smooth differential manifold that can be properly embedded into $\fpe$ via the embedding $\mathfrak{i}_{\elag_\m}$.
We take care of the validity of these assumptions case by case in the examples considered. 

We proceed along the section by recalling that the the differential form $\Pi_{\de \m}^\star \Omega^{\de \m}$ gives rise to a canonical structure on the space of solutions of the theory and that it is a pre-symplectic structure.

First of all, let us note that the differential form $\Pi^\star_{\de \m} \alpha^{\de \m}$ can be defined similarly for any codimension-$1$ hypersurface in $\m$, say $\Sigma$, in the following way:
\be
\pss \alpha^\Sigma_\chi (\mathbb{X}_\chi) \,=\, \int_{\Sigma} \chi^\star_\Sigma \left[\, i_X \dd \Theta_H \,\right] \, ,
\ee
where $\Pi_\Sigma$ is defined analogously to $\Pi_{\partial \mathscr{M}}$: $\Pi_{\Sigma} \;\; : \;\; \fpe \to \fpe^{\Sigma}$, $\chi \mapsto \chi\bigr|_{\Sigma}$.  
Its (minus) differential gives rise to the following $2$-form
\be\label{eq:preform}
\pssos_\chi(\mathbb{X}_\chi, \mathbb{Y}_\chi)\,=\, -\int_\Sigma \chi_\Sigma^\star \left[\, i_Y i_X \, \dd \Theta_H \,\right] \, ,
\ee
which is the analogue of $\Pi_{\de \m}^\star \Omega^{\de \m}$ for a generic co-dimension $1$ hypersurface in $\m$. 
We will say that a smooth hypersurface $\Sigma$ of the spacetime $\mathscr{M}$ is a slice of $\mathscr{M}$ if $\mathscr{M} \backslash \Sigma$ is the disjoint union of two spacetimes $\mathscr{M}_+$ and $\mathscr{M}_-$.
Now, the following, crucial, result holds \cite{Ciaglia-DC-Ibort-Marmo-Schiav-Zamp2021-Cov_brackets_toappear} 

\begin{proposition}
The differential 2-form $\pssos$ does not depend on the particular slice $\Sigma$ chosen if it is evaluated on the solutions of the equations of the motion.
\end{proposition}
An immediate consequence of this result is that the differential form $\mathfrak{i}_{\elag_\m}^\star \pssos$ does not depend on the particular slice $\Sigma$ chosen and, consequently, it defines a presymplectic structure on the Euler-Lagrange space  $\elag_\m$ because is closed by construction.
In particular when it has a non-empty kernel we will refer to the theory under investigation as a gauge theory.
That this notion coincides with the usual one in terms of principal fibre bundles is discussed in \cite{Ciaglia-DC-Ibort-Marmo-Schiav-Zamp2021-Cov_brackets_toappear}.

We conclude the section by discussing the relation between the presymplectic structure $\pssos$ and the presymplectic structure $\Omega_\infty^\Sigma$ which comes from the PCA analysis (\ref{eq:pca}).
Given a slice $\Sigma$, we will say that is defines a well posed boundary problem if the restriction of the map $\Pi_\Sigma$ to $\elag_\m$ is surjective, in other words, for any functions $(\varphi^a, p_a)$ on $\Sigma$ there is a solution $\chi\in \elag_\m$ possessing them as boundary data. 
First, the following result holds (see \cite{Ibort-Spivak2017-Covariant_Hamiltonian_YangMills, Asorey-Ibort-Spivak2017-Boundary_conditions}).
\begin{proposition}\label{prop:presymplectic}
Given a slice $\Sigma$, suppose that there exist an $\epsilon > 0$ such that there exists a collar $C^\Sigma_\epsilon \,=\, (-\epsilon,\, \epsilon) \times \Sigma$, such that $\Sigma$ defines a well posed boundary problem for the space of solutions $\elag_{C^\Sigma_\epsilon}$.  Then the space of solutions in such a collar $\elag_{C^\Sigma_\epsilon}$ consists of the solutions of the pre-symplectic Hamiltonian system $(\fpe^\Sigma,\, \os,\, \mathcal{H})$, where $\Omega^\Sigma$ is the differential form on $\fpe^\Sigma$ from which the structure $\pssos$ comes, Eq.  (\ref{eq:preform}), and $\mathcal{H}$ is the Hamiltonian functional:
\be \label{Eq:Hamiltonian boundary}
\mathcal{H} \,=\, \int_{\Sigma} \left(\beta^k_a \partial_k \varphi^a - H(\varphi,\, p,\, \beta)\right) vol_\Sigma
\ee
where $(\varphi^a,\, p_a,\, \beta_a^k) \,= \Pi_\Sigma (\phi^a,\, P^0_a,\, P^k_a) = (\phi^a,\, P^0_a,\, P^k_a)\bigr|_\Sigma$.
\end{proposition}
The form $\os$ is, in general, pre-symplectic and, thus, solutions of the pre-symplectic system above can be found via the pre-symplectic constraint algorithm, Eq. (\ref{eq:pca}). 
Denote by $\mathcal{M}_\infty$ the final stable manifold resulting from the pre-symplectic constraint algorithm applied to $(\fpe^\Sigma,\, \os,\, \mathcal{H})$.
Denote by $\mathfrak{i}_\infty$ its immersion into $\fpe^\Sigma$, by $\Omega^\Sigma_\infty \,=\, \mathfrak{i}_\infty^\star \Omega^\Sigma$, and by $\mathcal{H}_\infty \,=\, \mathfrak{i}_\infty^\star \mathcal{H}$. 
In case $\mathcal{M}_\infty$ is (strongly) symplectic, it can be proved \cite{Ciaglia-DC-Ibort-Marmo-Schiav-Zamp2021-Cov_brackets_toappear} that:
\be \label{Eq:relation pssos osinfty}
\pssos \,=\, \Psi^\star \os_\infty \, ,
\ee
where $\Psi =  F^{\mathbb{\Gamma}_\infty}  \circ\mathfrak{i}_\infty$, with $F^{\mathbb{\Gamma}_\infty}\colon (\varphi^a, p_a) \mapsto (\varphi^a_t , p_{a,t})$, being the flow of the Hamiltonian vector field associated to $\mathcal{H}_\infty$, which is a diffeomorphism under the conditions stated in the proposition.

When  $\mathcal{M}_\infty$ is a pre-symplectic manifold as well, it can be proved \cite{Ciaglia-DC-Ibort-Marmo-Schiav-Zamp2021-Cov_brackets_toappear}, that a relation of the type \eqref{Eq:relation pssos osinfty} holds, but only along a cross-section on the bundle $\mathcal{M}_\infty \to \mathcal{M}_\infty / K$ ($K$ denoting the characteristic distribution of $\os_\infty$) which represents the fixing of a particular gauge.


\section{Symmetry theory on the Euler-Lagrange space}
\label{Sec:Symmetry theory on the Euler-Lagrange space}

In this section we will apply the general theory developed in sections \ref{Sec:Symmetries for pre-symplectic dynamics} and \ref{Sec:The Euler-Lagrange space as pre-symplectic manifold} to the study of symmetry groups on the space-time underlying the field theory and investigate their properties. 
The corresponding momentum maps will be constructed for two instrumental examples, Klein-Gordon theory, Sect.  \ref{Subsec:The symplectic case: Klein-Gordon theory}, and Electrodynamics, Sect.  \ref{Subsec:Gauge theories: free Electrodynamics}, using the the coisotropic embedding in the presymplectic case.

\subsection{The symplectic case: Klein-Gordon theory}
\label{Subsec:The symplectic case: Klein-Gordon theory}

This section is devoted to apply the theory developed in Sect. \ref{Sec:Symmetries for pre-symplectic dynamics} to the particular case in which the symplectic manifold is the space of solutions of a first order Hamiltonian field theory without gauge symmetries.
We will consider the group of symmetries of the underlying space-time of the Klein-Gordon theory, consider its action upon the space of solutions of the theory and we will investigate whether it is canonical/Hamiltonian/strongly Hamiltonian.
Moreover, we will construct the related momentum map and conserved currents.

We consider the real scalar Klein-Gordon theory, representing an uncharged spinless massive boson field,  on the Minkowski space-time $(\mathbb{M},\, \eta)$ where $\eta$ is the Minkowski metric of signature $-+++$,  and where we chose the (global) chart $(U_{\mm},\,\psi_{U_{\mm}})$, $\psi_{U_{\mm}}(m)\,=\, (x^0,...,x^3) \,=:\, x$ where $m \in U_{\mm} \subset \mm$.
The bundle $\mathbb{E} \to \mm$ has typical fibre $\mathcal{E} \,=\, \mathbb{R}$ and, thus, the Covariant Phase Space reads $\pe \,=\, \mm \times \mathbb{R} \times \mathbb{R}^4$ where we chose the following adapted fibered chart $(U_{\pe},\, \psi_{U_{\pe}})$, $\psi_{U_{\pe}}(\mathfrak{p}) \,=\, (x^0,...,x^3,\, u,\, \rho^0,..., \rho^3)\,=:\,(x,\, u, \rho)$, where $\mathfrak{p} \in U_{\pe} \subset \pe$.
The Hamiltonian of the theory reads 
$$
H \,=\,  \frac{1}{2}(\eta_{\mu \nu} \rho^\mu \rho^\nu + m^2 u^2) \, ,
$$ 
with $\mu = 0, 1, 2, 3$, and the Hamiltonian functional \eqref{Eq:Hamiltonian boundary} becomes ($i,j = 1,2,3$):
$$
\mathcal{H} \,=\, \int_\Sigma \left[\, \beta^k \de \varphi - \frac{1}{2}\left( \delta_{jk}\beta^j \beta^k - p^2 - m^2 \varphi^2 \right) \,\right] vol_\Sigma \, ,
$$
for any $\Sigma$ a slice of $\mathbb{M}$ and $vol_\Sigma$ the volume form defined by the Riemannian metric induced on $\Sigma$ by $\eta$.
In particular $\Sigma$ will be chosen to be a Cauchy hypersurface, i.e., a space-like codimension 1 smooth submanifold that intersects each light ray in $\mathbb{M}$ at exactly one point.
The space of dynamical fields $\fpe$ is the space of sections of the bundle $\delta_1 \;\; : \;\; \pe \to \mm$ of the type 
\be
\chi \;\; : \;\; U_{\mm} \to \upe \;\; : \;\; \psi^{-1}_{U_{\mm}}(x) \mapsto \chi\left( \psi^{-1}_{U_{\mm}}(x) \right) \,=\, \left(\,\psi^{-1}_{U_\mathbb{I}}(x),\, \phi(\psi^{-1}_{U_\mathbb{I}}(x)),\, P^\mu(\psi^{-1}_{U_\mathbb{I}}(x)) \,\right)
\ee
where $\phi$ is a section of the bundle $\mathbb{E} \to \mm$ belonging to the Sobolev space $\mathcal{H}^1(\mm,\, vol_{\mm})$, i.e., the space of real-valued functions on $\mathbb{M}$ such that they are differentiable in the weak sense and their differentials are square integrable, and $P$ is a square integrable section of the pull-back bundle $\phi^\star\pe$.
This is consistent because the action functional originally defined on $\Gamma^\infty(\delta_1)$ can be extended by continuity to $\mathcal{H}^1(\mm,\, vol_{\mm}) \times \mathcal{L}^2(\phi^\star \pe)$ as discussed in more detail in \cite{Ciaglia-DC-Ibort-Marmo-Schiav-Zamp2021-Cov_brackets_toappear}.
For the sake of simplicity, we will simply denote them by $\chi(x) \,=\, (x,\, \phi(x),\,P^\mu(x))$.
The equations of motion read
\be
\frac{\de \phi}{\de x^\mu} \,=\, \eta_{\mu \nu} P^\nu \,, \qquad \, \frac{\de P^\mu}{\de x^\mu} \,=\, -m^2 \phi 
\ee
whose solutions are parametrized by the Cauchy data at $\Sigma$, $x^0 \,=\, x^0_\Sigma$, $(\varphi_\Sigma(\underline{x})\,=\, \phi(x^0\,=\, x^0_\Sigma,\underline{x}),\, p_\Sigma(\underline{x})\,=\, P^0(x^0\,=\, x^0_\Sigma,\, \underline{x})$ (where $\underline{x} \,=\, (x^1,\,x^2,\,x^3)$) in the following way
\be
\begin{split}
\elag_{\mm} = \biggl\{ &\chi \in \fpe \;\; : \\ &\phi(x^0,\,\underline{x}) = \int_{\mathbb{R}^3 \times \mathbb{R}^3} \left( \phi_\Sigma(\underline{x}') \mathrm{cos}\omega_k(x^0-x^0_\Sigma) - p_\Sigma(\underline{x}') \frac{\mathrm{sin}\omega_k(x^0-x^0_\Sigma)}{\omega_k} \right) e^{i \underline{k} \cdot (\underline{x}-\underline{x}')} \dd^3 k \dd^3 x' \\
&P^0(x^0,\,\underline{x}) = \int_{\mathbb{R}^3 \times \mathbb{R}^3} \left( \omega_k \phi_\Sigma(\underline{x}') \mathrm{sin}\omega_k(x^0-x^0_\Sigma) + p_\Sigma(\underline{x}') \mathrm{cos}\omega_k(x^0-x^0_\Sigma) \right) e^{i \underline{k} \cdot (\underline{x}-\underline{x}')} \dd^3 k \dd^3 x' \\
&P^k(x^0,\,\underline{x}) = \delta^{jk}\frac{\de \phi}{\de x^j}  \,\biggr\}
\end{split}
\ee
where $\omega_k \,=\, \sqrt{k^2 + m^2}$, $k^2 \,=\, k_1^2 + k_2^2 + k_3^2$.

The canonical structure of $\elag_{\mm}$ reads
\be
\pssos(\mathbb{X},\,\mathbb{Y})\,=\, \int_\Sigma \left[\, \mathbb{X}_u \mathbb{Y}_{\rho^0} - \mathbb{X}_{\rho^0} \mathbb{Y}_{u}  \,\right]_\Sigma vol_\Sigma
\ee
where $\mathbb{X},\, \mathbb{Y} \in \mathfrak{X} (\elag_{\mm})$ and $\mathbb{X}_u,\, \mathbb{X}_{\rho^0},\, \mathbb{Y}_u,\, \mathbb{Y}_{\rho^0}$ denote their components.
This structure is nondegenerate and, since the space of dynamical fields considered is a Hilbert manifold, it is strongly symplectic.

We consider the group of symmetries of the space-time $\mm$, that is, the Poincar\'e group $\mathscr{P} = \tau \rtimes \mathcal{L}$,
the semidirect product of the translations and the Lorentz group $\mathcal{L}$, contains the subgroup of translations, that we will denote by $\tau$, the subgroup of spatial rotations, that we will denote by $\mathscr{R}$ and the so called Lorentz boosts, that we will denote by $\mathscr{B}$.
The group we are dealing with is an affine group that can be represented via its natural action on $\mathbb{M}$ by means of $5 \times 5$ matrices
\be
\Phi_\tau^{\mm} \,=\, \left[\begin{matrix}
1 & 0 & 0 & 0 & a^0 \\
0 & 1 & 0 & 0 & a^1 \\
0 & 0 & 1 & 0 & a^2 \\
0 & 0 & 0 & 1 & a^3 \\
0 & 0 & 0 & 0 & 1
\end{matrix}  \right] \, ,
\ee
\be
\Phi_\mathscr{R}^{\mm} \,=\, \left[\begin{matrix}
1 & 0 & 0 & 0 & 0 \\
0 & \mathscr{R}_{11} & \mathscr{R}_{12} & \mathscr{R}_{13} & 0 \\
0 & \mathscr{R}_{21} & \mathscr{R}_{22} & \mathscr{R}_{23} & 0 \\
0 & \mathscr{R}_{31} & \mathscr{R}_{32} & \mathscr{R}_{33} & 0 \\
0 & 0 & 0 & 0 & 1
\end{matrix}  \right] \, ,
\ee
\be
\Phi_\mathscr{B}^{\mm} \,=\, \left[\begin{matrix}
\mathscr{B}_{00} & \mathscr{B}_{01} & \mathscr{B}_{02} & \mathscr{B}_{03} & 0 \\
\mathscr{B}_{10} & \mathscr{B}_{11} & \mathscr{B}_{12} & \mathscr{B}_{13} & 0 \\
\mathscr{B}_{20} & \mathscr{B}_{21} & \mathscr{B}_{22} & \mathscr{B}_{23} & 0 \\
\mathscr{B}_{30} & \mathscr{B}_{31} & \mathscr{B}_{32} & \mathscr{B}_{33} & 0 \\
0 & 0 & 0 & 0 & 1
\end{matrix}  \right] \, ,
\ee
obtained by taking the first $4$ components of the result of the action of the corresponding elements in the Poincar\'e group on the point with components $(x^0,\, x^1,\, x^2,\, x^3,\, 1)$
\be
\Phi_\tau^{\mm} \cdot m \,=\, \left[\begin{matrix}
1 & 0 & 0 & 0 & a^0 \\
0 & 1 & 0 & 0 & a^1 \\
0 & 0 & 1 & 0 & a^2 \\
0 & 0 & 0 & 1 & a^3 \\
0 & 0 & 0 & 0 & 1
\end{matrix}  \right] \left[\begin{matrix}
x^0 \\
x^1 \\
x^2 \\
x^3 \\
1
\end{matrix}  \right] \,=\, \left[\begin{matrix}
x^0 + a^0 \\
x^1 + a^1 \\
x^2 + a^2 \\
x^3 + a^3 \\
1
\end{matrix}  \right]
\ee
\be
\Phi_{\mathscr{R}}^{\mm} \cdot m \,=\, \left[ \begin{matrix}
1 & \boldsymbol{0} & 0 \\
0 & \mathscr{R}^j_k & \boldsymbol{0} \\
0 & \boldsymbol{0} & 1
\end{matrix} \right]\left[ \begin{matrix}
x^0 \\
x^k \\
1
\end{matrix} \right] \,=\, \left[\begin{matrix}
x^0 \\
\mathscr{R}^j_k x^k \\
1
\end{matrix}  \right]
\ee
where $\mathscr{R}^j_k$ are the components of a matrix in $O(3)$, and
\be
\Phi_{\mathscr{B}}^{\mm} \cdot m \,=\, \left[\begin{matrix}
\mathscr{B}^\mu_\nu & 0 \\
\boldsymbol{0} & 1
\end{matrix}  \right]\left[\begin{matrix}
x^\mu \\
1
\end{matrix} \right] \,=\, \left[\begin{matrix}
\mathscr{B}^\mu_\nu x^\nu \\
1
\end{matrix}  \right]
\ee
where $\mathscr{B}^\mu_\nu$ are the components of a matrix in $SO(1,3)$.
These actions can be lifted to a diffeomorphism of the bundle $\mathbb{E} \to \mm$, to $\pe$ (following the general theory of \cite{Asorey-Ibort-Spivak2017-Boundary_conditions}), and to $\elag_{\mm}$ in the following way 
\be
\Phi^{\pe}_\tau \;\; : \;\; \pe \to \pe \;\; :\;\; (x^\mu,\, u,\, \rho^\mu) \mapsto \Phi^{\pe}_\tau(x^\mu,\, u,\, \rho^\mu)\,=\,(x^\mu + a^\mu,\, u,\, \rho^\mu)
\ee
\be
\Phi^{\pe}_{\mathscr{R}} \;\; : \;\; \pe \to \pe \;\; :\;\; (x^\mu,\, u,\, \rho^\mu) \mapsto \Phi^{\pe}_{\mathscr{R}}(x^\mu,\, u,\, \rho^\mu) \,=\, \left(x^0,\, \mathscr{R}^j_k x^k,\, u,\, \rho^0,\, \mathscr{R}^j_k \rho^k \right)
\ee
\be
\Phi^{\pe}_{\mathscr{B}} \;\; : \;\; \pe \to \pe \;\; :\;\; (x^\mu,\, u,\, \rho^\mu) \mapsto \Phi^{\pe}_{\mathscr{B}}(x^\mu,\, u,\, \rho^\mu)\,=\,\left(\mathscr{B}^\mu_\nu x^\nu,\, u,\, \mathscr{B}^\mu_\nu \rho^\nu \right)
\ee
\be
\begin{split}
\Phi^{\elag_{\mm}}_\tau \;\; : \;\; \elag_{\mm} \to \elag_{\mm} \;\; :\;\; (x^\mu,\, \phi(x^\mu),\, P^\mu(x^\mu)) &\mapsto \Phi^{\elag_{\mm}}_\tau(x^\mu,\, \phi(x^\mu),\, P^\mu(x^\mu)) \,=\, \\ 
\,&=\,\left(x^\mu + a^\mu,\, \phi(x^\mu + a^\mu),\, P^\mu(x^\mu + a^\mu) \right)
\end{split}
\ee
\be
\begin{split}
\Phi^{\elag_{\mm}}_{\mathscr{R}} \;\; : \;\; \elag_{\mm} \to \elag_{\mm} \;\; &:\;\; (x^\mu,\, \phi(x^\mu),\, P^\mu(x^\mu)) \mapsto \Phi^{\elag_{\mm}}_{\mathscr{R}}(x^\mu,\, \phi(x^\mu),\, P^\mu(x^\mu)) \,=\, \\
\,&=\,\left(x^0,\, \mathscr{R}^j_k x^k,\, \phi(x^0,\, \mathscr{R}^j_k x^k),\, P^0(x^0,\, \mathscr{R}^j_k P^k(x^0,\,\mathscr{R}^j_k x^k),\, \mathscr{R}^j_k x^k) \right)
\end{split}
\ee
\be
\begin{split}
\Phi^{\elag_{\mm}}_{\mathscr{B}} \;\; : \;\; \elag_{\mm} \to \elag_{\mm} \;\; :\;\; (x^\mu,\, \phi(x^\mu),\, P^\mu(x^\mu)) & \mapsto \Phi^{\elag_{\mm}}_{\mathscr{B}}(x^\mu,\, \phi(x^\mu),\, P^\mu(x^\mu)) \,=\, \\
\,&=\,  (\mathscr{B}^\mu_\nu x^\nu,\, \phi(\mathscr{B}^\mu_\nu x^\nu),\, \mathscr{B}^\mu_\nu P^\nu(\mathscr{B}^\mu_\nu x^\nu)) \,.
\end{split}
\ee
The fact that this action sends solutions into solutions is readily proven by noting that $(\phi(x^0, x^j + a^j),\, P^\mu(x^0,\,x^j + a^j))$ is the solution associated with the Cauchy datum $(\varphi_\Sigma(\underline{x}-\underline{a}),\, p_\Sigma(\underline{x}-\underline{a}))$ at $x^0 \,=\, x^0_\Sigma$\red{\sout{,}} \blue{;} $(\phi(x^0 + a^0, x^j),\, P^\mu(x^0 + a^0,\,x^j))$ is the solution associated with the Cauchy datum $(\varphi_\Sigma(\underline{x}),\, p_\Sigma(\underline{x}))$ at $x^0 \,=\, x^0_\Sigma + a^0$ \red{\sout{,}} \blue{;} $(\phi(x^0 , \mathscr{R}^j_k x^k),\, P^\mu(x^0,\, \mathscr{R}^j_k x^k))$ is the solution associated with the Cauchy datum $(\varphi_\Sigma(\underline{x}),\, p_\Sigma(\underline{x}))$ at $x^0 \,=\, x^0_\Sigma$ and $(\phi(\mathscr{B}^\mu_\nu x^\nu),\, \mathscr{B}^\mu_\nu P^\nu(\mathscr{B}^\mu_\nu x^\nu))$ is the solution associated with the Cauchy datum $(\varphi_\Sigma(\mathscr{B}^j_\nu x^\nu),\, \mathscr{B}^\mu_\nu p_\Sigma^\nu(\mathscr{B}^j_\nu x^\nu))$ at $\mathscr{B}^0_\nu x^\nu \,=\, x^0_\Sigma$ where $p^\mu_\Sigma \,=\, (p_\Sigma,\, \delta^{jk} \de_j \phi_\Sigma)$.

The action of the Poincaré group onto $\elag_{\mm}$ written above is Hamiltonian for the symplectic structure $\pssos$.
Indeed, a function $J_\xi$ exists for all $\xi \in \mathfrak{g}$ such that
\be
\pssos(\mathbb{X}_\xi,\, \cdot \,) \,=\ J_\xi \,,
\ee
where $\mathbb{X}_\xi$ is the Killing vector field associated with $\xi$.

When $\xi \,=\, \xi_{\tau^0}$ is the element in the Lie algebra $\mathfrak{g} = \mathbb{R}^4 \rtimes \mathfrak{so}(1,3)$  of the Poincar\'e group $\mathscr{P}$, generating the translation $x^\mu \mapsto (x^0 + a^0,\, x^j)$, the corresponding Killing vector field reads
\be
\begin{split}
\mathbb{X}_{\xi_{\tau^0}} \,=\, \frac{d}{da^0} \left[\, \Phi_{e^{a^0 \xi_{\tau^0}}} \cdot \chi \,\right]_{a^0=0} \,&=\, \left(\, \left(\mathbb{X}_{\xi_{\tau^0}}\right)_u \,=\, \frac{\de \phi}{\de x^0},\, \left(\mathbb{X}_{\xi_{\tau^0}}\right)_{\rho^0}\,=\, \frac{\de P^0}{\de x^0},\, \left(\mathbb{X}_{\xi_{\tau^0}}\right)_{\rho^j} \,=\, \delta^{jk}\frac{\de}{\de x^k}\frac{\de \phi}{\de x^0} \,\right) \,=\, \\
\,&=\, \left(\, -P^0,\, -m^2 \phi - \Delta \phi,\, -\delta^{jk} \frac{\de P^0}{\de x^k} \,\right)
\end{split}
\ee
and, thus
\be
\pssos(\mathbb{X}_{\xi_{\tau^0}},\, \mathbb{Y}) \,=\, \int_\Sigma \left[\, -P^0 \mathbb{Y}_{\rho^0} + (m^2 \phi + \Delta \phi) \mathbb{Y}_u \,\right] vol_\Sigma \,.
\ee
On the other hand
\be
\dd J_{\xi_{\tau^0}}(\mathbb{Y}) \,=\, \int_{\mm} \left[\, \frac{\delta J_{\xi_{\tau^0}}}{\delta \phi} \mathbb{Y}_u + \frac{\delta J_{\xi_{\tau^0}}}{\delta P^0} \mathbb{Y}_{\rho^0} \,\right] \, vol_{\mm} \,.
\ee
Therefore, from the two latter equations one determines the function $J_{\xi_{\tau^0}}$ as the function (determined up to constants) satisfying, in the sense of distributions:
\be
\frac{\delta J_{\xi_{\tau^0}}}{\delta \phi} \delta(x^0-x^0_\Sigma) \,=\, (m^2 \phi + \Delta \phi) \,, \qquad \, \frac{\delta J_{\xi_{\tau^0}}}{\delta P^0} \delta(x^0-x^0_\Sigma) \,=\, -P^0 \,,
\ee
i.e.
\be
J_{\xi_{\tau^0}} \,=\, \int_\Sigma \frac{1}{2} \left[ m^2 \phi^2 - \delta^{jk}\frac{\de \phi}{\de x^j}\frac{\de \phi}{\de x^k} - {P^0}^2 \right]_\Sigma \, vol_\Sigma \,.
\ee
A similar computation can be performed for the elements $\xi_{\tau^k} \in \mathfrak{g}$ generating the translation $x^\mu \mapsto (x^0,\, x^j + \delta^{j}_k a^k)$, obtaining the function $J_{\xi_{\tau^k}}$ satisfying:
\be
\frac{\delta J_{\xi_{\tau^0}}}{\delta \phi} \delta(x^0-x^0_\Sigma) \,=\, -\frac{\de P^0}{\de x^k} \,, \qquad \, \frac{\delta J_{\xi_{\tau^0}}}{\delta P^0} \delta(x^0-x^0_\Sigma) \,=\, \frac{\de \phi}{\de x^k} \,,
\ee
i.e.
\be
J_{\xi_{\tau^k}} \,=\, \int_\Sigma P^0 \frac{\de \phi}{\de x^k}\biggr|_\Sigma vol_\Sigma  \,.
\ee
Note that $J_{\xi_{\tau^0}}$ and $J_{\xi_{\tau^k}}$ coincide with the charges associated with the energy-momentum tensor of the Klein-Gordon theory.

The momentum map is the map between $\elag_{\mm}$ and $\mathfrak{g}^\star$ satisfying
\be
\mathbb{J} \;\; : \;\; \elag_{\mm} \to \mathfrak{g}^\star \;\; :\;\; \langle\, \mathbb{J}(\chi) \,,\, \xi \,\rangle \,=\, J_\xi \,.
\ee
Let us compute it in the case $\xi \,=\, \xi_{\tau^0}$.
The element $\xi_{\tau^0} \in \mathfrak{g}$ such that $e^{a^0 \xi_{\tau^0}} \,=\, \Phi^{\mm}_{\tau^0}$ is represented as
\be
\Phi_{\xi_{\tau^0}} \,=\, \left[\begin{matrix}
0 & 0 & 0 & 0 & 1 \\
0 & 0 & 0 & 0 & 0 \\
0 & 0 & 0 & 0 & 0 \\
0 & 0 & 0 & 0 & 0 \\
0 & 0 & 0 & 0 & 0
\end{matrix}\right] \,.
\ee
Therefore, if we represent $\mathbb{J}(\chi)$ as
\be
\Phi_{\mathbb{J}} \,=\, \left[\begin{matrix}
\mathbb{J}_{00}(\chi) & \cdots & \mathbb{J}_{04}(\chi) \\
\vdots & & \vdots \\
\mathbb{J}_{40}(\chi) & \cdots & \mathbb{J}_{44}(\chi)
\end{matrix}\right]
\ee
we get 
\be
\langle \, \mathbb{J}(\chi) \,,\, \xi_{\tau^0} \, \rangle \,=\, \mathrm{Tr}\left[\, {\Phi_{\mathbb{J}}}^\dag \, \Phi^{\mm}_{\xi_{\tau^0}} \,\right] \,=\, \mathbb{J}_{04}(\chi)
\ee
and, consequently
\be
\mathbb{J}_{04}(\chi) \,=\, J_{\xi_{\tau^0}} \,.
\ee
A similar computation for $\xi \,=\, \xi_{\tau^k}$ gives
\be
\mathbb{J}_{0 (4-k)}(\chi) \,=\, J_{\xi_{\tau^k}} \,.
\ee
A computation analogous to the one made in the previous section shows that the momentum maps associated with $\xi_{\tau^0}$ and $\xi_{\tau^k}$ are equivariant.

Now, let us compute the conserved current and the momentum map in the case of a rotation $\mathscr{R}^{x^3}$ around the $x^3$ axis.
It is represented by the $5 \times 5$ matrix
\be 
\Phi^{\mm}_{\mathscr{R}^3} \,=\, \left[ \begin{matrix} 
1 & 0 & 0 & 0 & 0 \\
0 & \mathrm{cos}\theta & \mathrm{sin} \theta & 0 & 0 \\
0 & -\mathrm{sin} \theta & \mathrm{cos} \theta & 0 & 0 \\
0 & 0 & 0 & 1 & 0 \\
0 & 0 & 0 & 0 & 1
\end{matrix}\right] \,.
\ee
The Killing vector field associated with $\xi_{\mathscr{R}^3}$ reads
\be
\begin{split}
\mathbb{X}_{\xi_{\mathscr{R}^3}} \,=\, \frac{d}{d \theta}\left[\, \Phi^{\elag_{\mm}}_{e^{\theta \xi_{\mathscr{R}^3}}} \cdot \chi \,\right]_{\theta = 0} \,=\, \Biggl( \, & \left(\mathbb{X}_{\xi_{\mathscr{R}^3}}\right)_u \,=\, x^2 \frac{\de \phi}{\de x^1} - x^1 \frac{\de \phi}{\de x^2},\, \\ 
& \left(\mathbb{X}_{\xi_{\mathscr{R}^3}}\right)_{\rho^0} \,=\, x^2 \frac{\de P^0}{\de x^1} - x^1 \frac{\de P^0}{\de x^2},\, \\  
& \left(\mathbb{X}_{\xi_{\mathscr{R}^3}}\right)_{\rho^1} \,=\, x^2 \frac{\de P^1}{\de x^1} - x^1 \frac{\de P^1}{\de x^2} + P^2,\, \\ 
& \left(\mathbb{X}_{\xi_{\mathscr{R}^3}}\right)_{\rho^2} \,=\, x^2 \frac{\de P^2}{\de x^1} - x^1 \frac{\de P^2}{\de x^2} - P^1,\, \\ 
& \left(\mathbb{X}_{\xi_{\mathscr{R}^3}}\right)_{\rho^3} \,=\, x^2 \frac{\de P^3}{\de x^1} - x^1 \frac{\de P^3}{\de x^2} \, \Biggr) \,.
\end{split}
\ee
Consequently
\be
\pssos (\mathbb{X}_{\xi_{\mathscr{R}^3}},\, \mathbb{Y}) \,=\, \int_\Sigma \left[\, \left(x^2 \frac{\de \phi}{\de x^1} - x^1 \frac{\de \phi}{\de x^2} \right) \mathbb{Y}_{\rho^0} - \left(x^2 \frac{\de P^0}{\de x^1} - x^1 \frac{\de P^0}{\de x^2} \right) \mathbb{Y}_u \,\right] vol_\Sigma \,,
\ee
whereas 
\be
\dd J_{\xi_{\mathscr{R}^3}}(\mathbb{Y}) \,=\, \int_{\mm} \left[\, \frac{\delta J_{\xi_{\mathscr{R}^3}}}{\delta \phi} \mathbb{Y}_u + \frac{\delta J_{\xi_{\mathscr{R}^3}}}{\delta \phi} \mathbb{Y}_{\rho^0}  \,\right] vol_{\mm} \,.
\ee
Therefore $J_{\xi_{\mathscr{R}^3}}$ is the function satisfying, as always, in the sense of distributions:
\be
\begin{split}
\frac{\delta J_{\xi_{\mathscr{R}^3}}}{\delta \phi} \delta(x^0-x^0_\Sigma) \,=\, -x^2 \frac{\de P^0}{\de x^1} + x^1 \frac{\de P^0}{\de x^2}\, , \qquad \\   \frac{\delta J_{\xi_{\mathscr{R}^3}}}{\delta P^0}\delta(x^0-x^0_\Sigma) \,=\, x^2 \frac{\de \phi}{\de x^1} - x^1 \frac{\de \phi}{\de x^2}
\end{split}
\ee
i.e.
\be
J_{\xi_{\mathscr{R}^3}} \,=\, \int_\Sigma \left[\,P^0 \left(\,x^2 \frac{\de \phi}{\de x^1} - x^1 \frac{\de \phi}{\de x^2}\,\right)\right]_\Sigma vol_\Sigma \,.
\ee

An analogous calculation gives
\be
\begin{split}
J_{\xi_{\mathscr{R}^{x^2}}} \,=\, \int_\Sigma \left[\,P^0 \left(\,x^1 \frac{\de \phi}{\de x^3} - x^3 \frac{\de \phi}{\de x^1}\,\right)\right]_\Sigma vol_\Sigma \,, \qquad \\  J_{\xi_{\mathscr{R}^{x^1}}} \,=\, \int_\Sigma \left[\, P^0 \left(\,x^2 \frac{\de \phi}{\de x^3} - x^3 \frac{\de \phi}{\de x^2}\,\right)\right]_\Sigma vol_\Sigma \,.
\end{split}
\ee
Note that $J_{\xi_{\mathscr{R}^{x^j}}}$ are the $j$-components ($j$ going from $1$ to $3$) of the divergence of the generalized angular momentum of the Klein-Gordon field.

Now, to compute the momentum map associated with $\xi_{\mathscr{R}^{x^k}}$ one can recall that the group of spatial rotations is the subgroup $SO(3)$ of the Poincarè group and that its Lie algebra is isomorphic with $\mathbb{R}^3$.
Consequently, also the dual of $\mathfrak{so}(3)$ is isomorphic with $\mathbb{R}^3$ and, thus, if we consider $\xi_{\mathscr{R}^{x^j}}$ as a basis for $\mathfrak{so}(3)$, the components of the momentum map satisfying $\langle \mathbb{J}(\chi),\, \xi_{\mathscr{R}^{x^j}} \rangle \,=\, J_{\xi_{\mathscr{R}^{x^j}}} $ read
\be
\mathbb{J}(\chi)_j \,=\, J_{\xi_{\mathscr{R}^{x^j}}} \;\;\; j \,=\, 1, 2, 3 \,.
\ee

The computations of the currents associated to the boosts of the Poincar\'e group are analogous to the ones we did for the rotation if one recalls that the boosts along the three axis can be represented as
\be
\begin{split}
\Phi^{\mm}_{\mathscr{B}^{x^1}} &\,=\, \left[ \begin{matrix} 
\mathrm{cosh}w & -\mathrm{sinh} w & 0 & 0 & 0 \\
-\mathrm{sinh} w & \mathrm{cosh}w & 0 & 0 & 0 \\
0 & 0 & 1 & 0 & 0 \\
0 & 0 & 0 & 1 & 0 \\
0 & 0 & 0 & 0 & 1
\end{matrix}\right]  \\ 
\Phi^{\mm}_{\mathscr{B}^{x^2}} &\,=\, \left[ \begin{matrix} 
\mathrm{cosh}w & 0 & -\mathrm{sinh} w & 0 & 0 \\
0 & 1 & 0 & 0 & 0 \\
-\mathrm{sinh} w & 0 &  \mathrm{cosh}w & 0 & 0 \\
0 & 0 & 0 & 1 & 0 \\
0 & 0 & 0 & 0 & 1
\end{matrix}\right]  \\ 
\Phi^{\mm}_{\mathscr{B}^{x^3}} &\,=\, \left[ \begin{matrix} 
\mathrm{cosh}w & 0 & 0 & -\mathrm{sinh} w & 0 \\
0 & 1 & 0 & 0 & 0 \\
0 & 0 &  1 & 0 & 0 \\
-\mathrm{sinh} w & 0 & 0 & \mathrm{cosh}w & 0 \\
0 & 0 & 0 & 0 & 1
\end{matrix} \right]  \, ,
\end{split}
\ee
where $w$ represents the rapidity of the boost.
The three currents obtained in this case are
\be
\begin{split}
J_{\xi_{\mathscr{B}^{x^1}}} &\,=\, - \int_\Sigma \left[\, P^0\left(\, x^1 \frac{\de \phi}{\de x^0} + x^0 \frac{\de \phi}{\de x^1} \,\right) + P^1 \phi \, \right]_\Sigma vol_\Sigma \\
J_{\xi_{\mathscr{B}^{x^2}}} &\,=\, -\int_\Sigma \left[\, P^0\left(\, x^2 \frac{\de \phi}{\de x^0} + x^0 \frac{\de \phi}{\de x^2} \,\right) + P^2 \phi \,\right]_\Sigma vol_\Sigma \\
J_{\xi_{\mathscr{B}^{x^3}}} &\,=\, - \int_\Sigma \left[\, P^0\left(\, x^3 \frac{\de \phi}{\de x^0} + x^0 \frac{\de \phi}{\de x^3} \,\right) + P^3 \phi \,\right]_\Sigma vol_\Sigma
\end{split}
\ee
which are the remaining three independent components of the divergence of the generalized angular momentum usually found in Field Theory textbooks.
The components of the momentum map in this case are
\be
\mathbb{J}(\chi)_j \,=\,  J_{\xi_{\mathscr{B}^{x^j}}}  \;\;\; j \,=\, 1, 2, 3 \,.
\ee


\subsection{Gauge theories: free Electrodynamics}
\label{Subsec:Gauge theories: free Electrodynamics}

In this section we will apply the theory developed in Sect. \ref{Sec:Symmetries for pre-symplectic dynamics} to the case in which the pre-symplectic manifold is the space of solutions of a gauge theory.

In the example considered, the underlying space-time is, as in the Klein-Gordon theory discussed in Sect. \ref{Subsec:The symplectic case: Klein-Gordon theory}, the Minkowski space-time $(\mm,\, \eta)$ where we chose the (global) chart $(U_{\mm},\, \psi_{U_{\mm}})$, $\psi_{U_{\mm}}(m) \,=\, x^\mu$ where $m \in U_{\mm} \subset \mm$.
The bundle $\mathbb{E} \to \mm$ is the cotangent bundle over $\mm$, i.e., the typical fibre is $\mathcal{E} \,=\, \mathbf{T}^\star_m \mm$ and the configuration fields $\phi$ are (Lie algebra-valued\footnote{In this case the Lie algebra is that of the abelian group $U(1)$ which is isomorphic with $\mathbb{R}$.}) one-forms over $\mm$ that we will denote by $\phi \,=\, A \,=\, A_\mu(x) \dd x^\mu$.

The covariant phase space  $\pe$ has typical fibre $\mathbf{T}^\star_m \mm \otimes \left(\, \bigwedge^2 \mathbf{T}_m\mm \,\right)$.
We chose on it the following adapted fibered chart $(U_{\pe},\, \psi_{\upe})$, $\psi_{\upe}(\mathfrak{p}) \,=\, (u_\mu,\, \rho^{\mu \nu},\, x^\mu)$ where $\mathfrak{p} \in \upe \subset \pe$.
The space of dynamical fields $\fpe$ is the space of factorizable sections of the bundle $\delta_1 \;\; :\;\; \pe \to \mm$, (Eq. \eqref{eq:factorization}):
\be
\chi \;\; : \;\; U_{\mm} \to \upe \;\; :\;\; \psi^{-1}_{\mm}(x^\mu) \mapsto \chi\left(\psi^{-1}_{U_{\mm}}(x^\mu)\right) \,=\, \left(\,\psi^{-1}_{U_{\mm}}(x^\mu),\,  A_\mu(\psi^{-1}_{U_{\mm}}(x^\mu)),\, P^{\mu \nu}(\psi^{-1}_{U_{\mm}}(x^\mu)) \,\right)
\ee
where $A_\mu$ are sections of $\mathbb{E} \to \mm$ belonging to the Sobolev space $\mathcal{H}^1(\mm,\, vol_{\mm})$ and $P^{\mu \nu}$ are square integrable sections of the pull-back bundle $A^\star \pe$.
This is consistent  for the same reasons of the previous example.
For the sake of simplicity, we will denote them by: 
$$
\chi(x^\mu) \,=\, (x^\mu,\, A_\mu(x^\nu),\, P^{\mu \nu}(x^\sigma))\, .
$$
The Hamiltonian of the theory is just the quadratic function:
$$
H \,=\, \frac{1}{2} \eta_{\mu \alpha} \eta_{\nu \beta} \rho^{\mu \nu} \rho^{\alpha \beta} \, ,
$$ 
and the Hamiltonian functional is 
\be
\mathcal{H} \,=\, \int_{\Sigma}\left[ p^k \partial_k a_{0} + \frac{1}{2} \beta^{kj} \left( \partial_k a_{j} - \partial_j a_{k} \right) + \frac{1}{4} \delta_{jl}\delta_{km}\beta^{jk}\beta^{lm} + \frac{1}{2} \delta_{jk} p^k p^{j} \right]\mathrm{vol}_{\Sigma} 
\ee
where $\Sigma$ is any slice of $\mathbb{M}$, that for definiteness we take to be a Cauchy hypersurface diffeomorphic to $\mathbb{R}^3$ defined by $x^0 \,=\, x^0_\Sigma$, and $a_\mu$, $p^k$ and $\beta^{jk}$ denote the restrictions of $A_\mu$, $P^{0k}$ and $P^{jk}$ respectively, to $\Sigma$.
The equations of the motion are: 
\be\begin{split}
F_{\mu \nu} + \frac{1}{2} \eta_{\mu \alpha} \eta_{\nu \beta} P^{\alpha \beta} \,&=\, 0 \,, \\ \frac{\partial P^{\mu \nu}}{\de x^\mu} \,&=\, 0 
\end{split}
\ee
where $F_{\mu \nu} \,=\, \frac{1}{2}(\de_\mu A_\nu - \de_\nu A_\mu)$, and whose solutions are given by:
\be
\begin{split}
	A_{k}(x^\mu) \,=\, \frac{1}{4 \pi} \biggl[\, &\int_{\Sigma} \left(\, a^\Sigma_{k}(\underline{y}) + |x^0 - x^0_\Sigma| \delta_{kj}p^j_\Sigma(\underline{y}) \,\right) \tilde{G}_{\underline{x}, \, |x^0-x^0_\Sigma|}(\underline{y}) \, vol_\Sigma \, + \\ 
	&\int_{\Sigma}|x^0 - x^0_\Sigma| a^\Sigma_k(\underline{y}) \frac{\partial}{\partial s}\tilde{G}_{\underline{x}, (x^0- x^0_\Sigma)}(\underline{y}) (\Theta(x^0-x^0_\Sigma)-\Theta(x^0_\Sigma-x^0)) \, vol_\Sigma  \biggr]
\end{split}
\ee
where the integration on $\Sigma$ is over the $\underline{y}$ variables and $\tilde{G}_{\underline{x}, a}(\underline{y})$ is the characteristic function of the surface of the sphere with center $\underline{x}$ and radius $a$ and $\Theta$ is the Heaviside function, i.e., the discontinuous function being $1$ for its argument being positive and $0$ for its argument being less then or equal to $0$.
On the other hand $A_0$ is completely undetermined by the equations of the motion and the standard treatment is to fix it using a suitable gauge fixing (in \cite{Ciaglia-DC-Ibort-Marmo-Schiav-Zamp2021-Cov_brackets_toappear} the Coulomb gauge is fixed) while the $P^{\mu \nu}$ are determined by the constraint equation 
\be
F_{\mu \nu} + \frac{1}{2} \eta_{\mu \alpha} \eta_{\nu \beta} P^{\alpha \beta} \,=\, 0 \,.
\ee

As it is proven in \cite{Gotay1979-PhD_thesis} (see, for instance, \cite{Ibort-Spivak2017-Covariant_Hamiltonian_YangMills} and references therein for the geometric treatment of the non-Abelian case), the final manifold of constraints obtained by using the PCA, $\mathcal{M}_\infty$, is:
\be
\mathcal{M}_\infty \,=\, \left\{\, (a_k,\, p^k) \;\; : \;\;  \de_k p^k \,=\, 0 \;\; k \,=\, 1, 2, 3 \,\right\} \,=\, \mathcal{H}^1(\Sigma) \times \mathcal{H}(\mathrm{div}0, \Sigma)
\ee
where $\mathcal{H}(\mathrm{div}0, \Sigma)$ denotes the closed Hilbert subspace of $\mathcal{L}^2(\Sigma, vol_\Sigma)$ such that $\de_k p^k \,=\, 0$ (see \cite[Eq. (1.41), Eq. (1.42)]{Lions1990-Spectral_theory}).
The structure $\os_\infty$ reads
\be
\os_\infty(\mathbb{X},\, \mathbb{Y}) \,=\, \int_\Sigma \left( \mathbb{X}_{a_k} \mathbb{Y}_p^k - \mathbb{X}_p^k\, \mathbb{Y}_{a_k} \right) \, vol_\Sigma \,.
\ee
The kernel of $\os_\infty$ consists of vector fields with $\mathbb{X}_{a_k} \,=\, \de_k \phi$ for some $\phi$.
They generate (via their flow) the following action on $\mathcal{M}_\infty$
\be
\Phi_{\mathbb{X}_{\de \phi}} \;\; : \;\; \mathcal{M}_\infty \to \mathcal{M}_\infty \;\; : \;\; m_\infty \mapsto \Phi_{\mathbb{X}_{\de \phi}} \cdot m_\infty \,=\, (a_k + \de_k \phi,\, p^k) \,.
\ee
In order for this action to be well defined we choose the space of gauge functions $\phi$ to belong to $\mathcal{H}^2(\Sigma)$.
Now, let us note that the following splitting of $\mathcal{M}_\infty$ exists
\be
\mathcal{M}_\infty \,\simeq\, \mathcal{H}^1(\mathrm{div}0, \Sigma) \times \mathrm{grad} \mathcal{H}^2(\Sigma) \times \mathcal{H}(\mathrm{div}0, \Sigma) 
\ee
which comes from the the following decomposition of $\mathcal{H}^1(\Sigma)$ \cite{Lions1990-Spectral_theory}
\be 
\mathcal{H}^1(\Sigma) \,=\, \mathcal{H}^1(\mathrm{div}0, \Sigma) \oplus \mathrm{grad} \mathcal{H}^2(\Sigma) \,,
\ee
where $\mathrm{grad }\mathcal{H}^2(\Sigma)$ is the image of the $\mathrm{grad}$ operator acting on $\mathcal{H}^2(\Sigma)$.
Accordingly with the latter decomposition we will denote points in $\mathcal{M}_\infty$ by $m_\infty \,=\, (\tilde{a}_k,\, \partial_k \phi,\, p^k)$, where $\tilde{a}_k \in \mathcal{H}^1(\mathrm{div}0, \Sigma)$.
By virtue of the decomposition above, the tangent space of $\mathcal{M}_\infty$ at some point $m_\infty$ also splits as
\be \label{Eq: decomposition tangent space minfty electrodynamics}
\mathbf{T}_{m_\infty} \mathcal{M}_\infty \,=\, \mathcal{H}^1(\mathrm{div}0, \Sigma) \times \mathrm{grad} \mathcal{H}^2(\Sigma) \times \mathcal{H}(\mathrm{div}0, \Sigma) \, ,
\ee
because of the natural identifications: $\mathbf{T}_{m_\infty} \mathcal{H}^1(\mathrm{div}0, \Sigma) \cong \mathcal{H}^1(\mathrm{div}0, \Sigma)$, $\mathbf{T}_{m_\infty} \mathrm{grad}\mathcal{H}^2(\Sigma)  \cong  \mathrm{grad}\mathcal{H}^2(\Sigma) $ and $\mathbf{T}_{m_\infty} \mathcal{H}(\mathrm{div}0, \Sigma) \cong \mathcal{H}(\mathrm{div}0, \Sigma)$.
The second term in the right hand side of (\ref{Eq: decomposition tangent space minfty electrodynamics}) clearly represents the kernel of $\os_\infty$, say $K$.
A connection on the bundle $\mathcal{M}_\infty \to \mathcal{M}_\infty / \mathrm{grad}\mathcal{H}^2(\Sigma)$ exists whose horizontal vector fields lie, at each point, in the complement of $K$ in the decomposition \eqref{Eq: decomposition tangent space minfty electrodynamics} and is the so called Coulomb connection
\be\label{eq:coulomb}
\mathcal{R} \;\; : \;\; \mathfrak{X}(\mathcal{M}_\infty) \to \mathfrak{X}^{H}(\mathcal{M}_\infty)\,=\,\mathcal{H}^1(\mathrm{div}0,\, \Sigma)\times \mathcal{H}(\mathrm{div}0, \Sigma)  \;\; : \;\; \mathbb{X} \mapsto \mathcal{R}(\mathbb{X}) \,=\, \tilde{\mathbb{X}}
\ee
where the tangent vector $\tilde{\mathbb{X}}$ is given by:
\be
\tilde{\mathbb{X}} \,=\, \left(\,\tilde{\mathbb{X}}_{a_k} \,=\, \mathbb{X}_{a_k} - \de_k \int_{\Sigma} G_\Delta \delta^{jl}\de_j \mathbb{X}_{a_l} \, vol_\Sigma,\, \tilde{\mathbb{X}}_p^k \,=\,\mathbb{X}_p^k \, \right)
\ee
and $G_\Delta$ represents the Green function of the Laplacian operator.

The coisotropic embedding theorem in this case leads to the following symplectic manifold:
\be
\begin{split}
\mathcal{P} \,&=\, \mathcal{H}^1(\mathrm{div}0,\Sigma) \times \mathrm{grad}\mathcal{H}^2(\Sigma) \times \mathcal{H}(\mathrm{div}0, \Sigma) \times K^\star \,\simeq \, \\
\,&\simeq \, \mathcal{H}^1(\mathrm{div}0,\Sigma) \times \mathrm{grad}\mathcal{H}^2(\Sigma) \times \mathcal{H}(\mathrm{div}0, \Sigma) \times \mathrm{grad}\mathcal{H}^2(\Sigma)^\star \, ,
\end{split}
\ee
We denote points in $\mathcal{P}$ by $\tilde{m}_\infty \,=\, (\tilde{a}_k,\, \de_k \phi,\, p^k,\, \mu^k)$ and the symplectic structure $\Omega^\mathcal{P}$ reads:
\be
\Omega^{\mathcal{P}}(\mathbb{X}\otimes \kappa,\, \mathbb{Y} \otimes \kappa') \,=\, \int_\Sigma \left( {\mathbb{X}_{\tilde{a}}}_k \mathbb{Y}_p^k - \mathbb{X}_p^k {\mathbb{Y}_{\tilde{a}}}_k \right) vol_\Sigma + \langle \, \kappa,\,  {\mathbb{Y}_{\de \phi}} \, \rangle - \langle \,\kappa',\, {\mathbb{X}_{\de \phi}} \, \rangle \,,
\ee
where $\langle \,\cdot \,,\, \cdot \,\rangle$ represents the pairing between $\mathrm{grad}\mathcal{H}^2(\Sigma)$ and its dual.
Now, let us consider the group of symmetries of the space-time $\mm$, i.e., the Poincar\'e group.
In particular, let us focus on the group of spatial rotations. 
They act on $\mm$ by means of $SO(3)$ matrices
\be
\Phi_{\mathscr{R}}^{\mm} \;\; : \;\; \mm \to \mm \;\; :\;\;  x^k \mapsto \Phi_{\mathscr{R}}^{\mm}(x)^k \,=\, \mathscr{R}^k_j x^j
\ee
where $\mathscr{R}^k_j$ represent the matrix elements of the $SO(3)$ matrix $\Phi_\mathscr{R}^{\mm}$.
Since the fields $a$ are $1$-forms on $\mm$ and the fields $p$ are contravariant tensors on $\mm$, they transform via the pull-back (via $\Phi_{\mathscr{R}^{-1}}^{\mm}$) and via the push-forward (via $\Phi_\mathscr{R}^{\mm}$) respectively.
Therefore, the action of $\mathscr{R}$ is lifted to $\mathcal{M}_\infty$ as: 
\be
\begin{split}
\Phi_{\mathscr{R}}^{\mathcal{M}_\infty} \;\; : \;\; \mathcal{M}_\infty \to \mathcal{M}_\infty  \;\; :\;\; &(\tilde{a}_k(x),\, \de_k \phi(x),\, p^k(x)) \mapsto \\
 \mapsto &\left( {\mathscr{R}^{-1}}^j_k \tilde{a}_j\left(\Phi^{\mm}_\mathscr{R} \cdot x \right),\, {\mathscr{R}^{-1}}^j_k \de_j \phi \left(\Phi^{\mm}_\mathscr{R} \cdot x \right),\, \mathscr{R}^k_j p^j(\Phi_{\mathscr{R}}^{\mm} \cdot x)  \right) \,.
\end{split}
\ee
This action can be also lifted to $\mathcal{P}$ via the procedure explained in Sect. \ref{Sec:Symmetries for pre-symplectic dynamics} by computing the dual action of its tangent lift.
A straightforward calculation gives the following action on $\mathcal{P}$:
\be \label{Eq: action coisotropic embedding electrodynamics}
\begin{split}
\Phi^{\mathcal{P}}_{\mathscr{R}} \;\; : \;\; \mathcal{P} \to \mathcal{P} \;\; : \;\; &\left( \tilde{a}_k(x),\, \de_k \phi(x),\, p^k(x),\, \mu^k(x)  \right) \mapsto \\
\mapsto &\left( {\mathscr{R}^{-1}}^j_k \tilde{a}_j\left(\Phi^{\mm}_\mathscr{R} \cdot x \right),\, {\mathscr{R}^{-1}}^j_k \de_j \phi \left(\Phi^{\mm}_\mathscr{R} \cdot x \right),\, \mathscr{R}^k_j p^j(\Phi_{\mathscr{R}}^{\mm} \cdot x),\, \mathscr{R}^k_j \mu^j \left( \Phi^{\mm}_\mathscr{R} \cdot x \right)  \right) \,.
\end{split}
\ee
Let us focus for a moment in the case in which $\mathscr{R}$ is a rotation around the $x^3$-axis, denoted, for short, $\mathscr{R}^3$.
Now, we prove that the action above is canonical and give rise to a momentum map satisfying
\be \label{Eq: hamiltonian vf coisotropic embedding electrodynamics}
\Omega^\mathcal{P}(\mathbb{X}_{\xi_{\mathscr{R}^3}},\, \mathbb{Y} \oplus \kappa') \,=\, \dd J_{\xi_{\mathscr{R}^3}}(\mathbb{Y} \oplus \kappa')
\ee
where $\mathbb{X}_{\xi_{\mathscr{R}^3}}$ is the generator of the action \eqref{Eq: action coisotropic embedding electrodynamics}.
In the case considered $\xi_{\mathscr{R}^3}$ is an element in $\mathfrak{so}(3)$ represented by the matrix:
\be
\Phi_{\xi_{\mathscr{R}^3}} \,=\, \left[ \begin{matrix}
0 & 0 & -1 \\
0 & 0 & 0 \\
1 & 0 & 0
\end{matrix} \right] \, ,
\ee
and the vector field $\mathbb{X}_{\xi_{\mathscr{R}^3}}$ is: 
\be 
\begin{split}
\mathbb{X}_{\xi_{\mathscr{R}^3}} \,=\, \frac{d}{d\theta} \left[ \Phi^{\mathcal{P}}_{e^{\theta \xi_{\mathscr{R}^3}}} \cdot \tilde{m}_\infty \right]_{\theta \,=\, 0} \,=\, \left(\, \left(\mathbb{X}_{\xi_{\mathscr{R}^3}}\right)_{\tilde{a}_k},\, \left(\mathbb{X}_{\xi_{\mathscr{R}^3}}\right)_{\de\phi_k},\, \left(\mathbb{X}_{\xi_{\mathscr{R}^3}}\right)_p^k,\, \left(\mathbb{X}_{\xi_{\mathscr{R}^3}}\right)_\mu^k  \,\right)
\end{split}
\ee
where
\be
\left(\mathbb{X}_{\xi_{\mathscr{R}^3}}\right)_{\tilde{a}_k} \,=\, \left[\begin{matrix}
(\ltre)\tilde{a}_1 + \tilde{a}_2 \\
-\tilde{a}_1 + (\ltre)\tilde{a}_2 \\
(\ltre)\tilde{a}_3 
\end{matrix}\right] \ee \be \left(\mathbb{X}_{\xi_{\mathscr{R}^3}}\right)_{\de \phi_k} \,=\, \left[\begin{matrix}
(\ltre)\de_1 \phi + \de_2 \phi \\
-\de_1 \phi + (\ltre)\de_2 \phi \\
(\ltre)\de_3 \phi 
\end{matrix}\right]
\ee
\be
\left(\mathbb{X}_{\xi_{\mathscr{R}^3}}\right)_p^k \,=\, \left[\begin{matrix}
(\ltre)p^1 + p^2 \\
-p^1 + (\ltre)p^2 \\
(\ltre)p^3 
\end{matrix}\right] \ee \be \left(\mathbb{X}_{\xi_{\mathscr{R}^3}}\right)_\mu^k \,=\, \left[\begin{matrix}
(\ltre)\mu^1 + \mu^2 \\
-\mu^1 + (\ltre)\mu^2 \\
(\ltre)\mu^3 
\end{matrix}\right]
\ee
Therefore, the left hand side of \eqref{Eq: hamiltonian vf coisotropic embedding electrodynamics} is
\be
\int_\Sigma \left( \left( \mathbb{X}_{\xi_{\mathscr{R}^3}}\right)_{\tilde{a}_k} \mathbb{Y}_p^k - \left(\mathbb{X}_{\xi_{\mathscr{R}^3}}\right)_p^k {\mathbb{Y}_{\tilde{a}}}_k \right) vol_\Sigma + \langle \, \left(\mathbb{X}_{\xi_{\mathscr{R}^3}}\right)_\mu ,\, {\mathbb{Y}_{\de \phi}}_k \, \rangle - \langle \, \kappa' ,\, \left(\mathbb{X}_{\xi_{\mathscr{R}^3}}\right)_{\de \phi} \, \rangle \,.
\ee
On the other hand, the right hand side of \eqref{Eq: hamiltonian vf coisotropic embedding electrodynamics} reads
\be
\int_\Sigma \left( \frac{\delta J_{\xi_{\mathscr{R}^3}}}{\delta \tilde{a}_k} {\mathbb{Y}_{\tilde{a}}}_k + \frac{\delta J_{\xi_{\mathscr{R}^3}}}{\delta \de_k \phi} {\mathbb{Y}_{\de \phi}}_k + \frac{\delta J_{\xi_{\mathscr{R}^3}}}{\delta p^k} \mathbb{Y}_p^k \right) vol_\Sigma + \langle\, \frac{\delta J_{\xi_{\mathscr{R}^3}}}{\delta \mu} \,,\,\, \kappa' \rangle \,,
\ee
which gives
\be
\frac{\delta J_{\xi_{\mathscr{R}^3}}}{\delta \tilde{a}_k} \,=\, \left[\begin{matrix}
-(\ltre)p^1 - p^2 \\
p^1 - (\ltre)p^2 \\
-(\ltre)p^3 
\end{matrix}\right] \ee\be \frac{\delta J_{\xi_{\mathscr{R}^3}}}{\delta \de_k \phi} \,=\, \left[\begin{matrix}
(\ltre)\mu^1 + \mu^2 \\
-\mu^1 + (\ltre)\mu^2 \\
(\ltre)\mu^3 
\end{matrix}\right]
\ee
\be
\frac{\delta J_{\xi_{\mathscr{R}^3}}}{\delta p^k} \,=\, \left[\begin{matrix}
(\ltre)\tilde{a}_1 + \tilde{a}_2 \\
-\tilde{a}_1 + (\ltre)\tilde{a}_2 \\
(\ltre)\tilde{a}_3 
\end{matrix}\right] \ee\be \frac{\delta J_{\xi_{\mathscr{R}^3}}}{\delta \mu^k} \,=\, \left[\begin{matrix}
-(\ltre)\de_1 \phi - \de_2 \phi \\
\de_1 \phi - (\ltre)\de_2 \phi \\
-(\ltre)\de_3 \phi 
\end{matrix}\right]
\ee
and, thus
\be \label{Eq: momentum map electrodynamics}
\begin{split}
 & J_{\xi_{\mathscr{R}^3}} (\tilde{a}_k,\, \de_k \phi,\, p^k,\, \mu^k) \,=\, \\ & = \int_{\Sigma} \left[ (\ltre) \tilde{a}_1 p^1 + \tilde{a}_2 p^1 + (\ltre)\tilde{a}_2 p^2 - \tilde{a}_1 p^2 + (\ltre) \tilde{a}_3 p^3 \right] vol_\Sigma \, \\ 
&+  \langle \mu^1 \,, (\ltre) \de_1 \phi - \de_2 \phi \rangle + \langle \mu^2 \,,(\ltre) \de_2 \phi + \de_1 \phi \rangle + \langle  \mu^3 \,,(\ltre) \de_3 \phi \rangle \,.
\end{split}
\ee
Let us note that the quantities in the right arguments of the pairings above are the components of the vector field $\mathbb{X}_{\xi_{\mathscr{R}^3}}$ that correspond to the components of the vector field $\mathbb{X}_{\xi_{\mathscr{R}^3}}^H$, being the horizontal lifting of the vector field $\mathbb{X}_{\xi_{\mathscr{R}^3}}$ with respect to the connection $\mathcal{R}^\uparrow$ that extends the Coulomb connection $\mathcal{R}$ introduced above, Eq. (\ref{eq:coulomb}):
\be
\mathcal{R}^{\uparrow} \;\; : \;\; \mathfrak{X}(\mathcal{P}) \to \mathfrak{X}^H(\mathcal{P}) \;\; :\;\; \mathbb{X}\oplus \kappa \mapsto \mathcal{R}^\uparrow(\mathbb{X}) \,=\, \mathbb{X}^H = \tilde{\mathbb{X}}\oplus \tilde{\kappa}
\ee
where $\tilde{\mathbb{X}} \,=\, \mathcal{R}(\mathbb{X})$ and $\tilde{\kappa} \,=\, \kappa$.
With this in mind the function \eqref{Eq: momentum map electrodynamics} has exactly the expression \eqref{Eq: momentum map coisotropic embedding} for the connection $A \,=\, \mathcal{R}^\uparrow$.

The currents $J_{\xi_{\mathscr{R}^{x^1}}}$ and $J_{\xi_{\mathscr{R}^{x^2}}}$ can be computed in the same way obtaining
\be
\begin{split}
J_{\xi_{\mathscr{R}^{x^1}}}& \,=\, \int_{\Sigma} \left[ (\ldue) \tilde{a}_1 p^1  + (\ldue)\tilde{a}_2 p^2 + \tilde{a}_3 p^2 + (\ldue) \tilde{a}_3 p^3 - \tilde{a}_2 p^3 \right] vol_\Sigma + \\ 
+& \langle \mu^1, (\ldue) \de_1 \phi  \rangle + \langle \mu^2 ,(\ldue) \de_2 \phi - \de_3 \phi \rangle + \langle \mu^3, (\ldue) \de_3 \phi + \de_2 \phi \rangle\, , \nonumber
\end{split}
\ee
and
\be
\begin{split}
J_{\xi_{\mathscr{R}^{x^2}}}& \,=\, \int_{\Sigma} \left[ (\luno) \tilde{a}_1 p^1 + \tilde{a}_3 p^1 + (\luno)\tilde{a}_2 p^2  + (\luno) \tilde{a}_3 p^3  - \tilde{a}_1 p^3 \right] vol_\Sigma + \\ 
+& \langle \mu^1 \,, (\luno) \de_1 \phi - \de_3 \phi \rangle + \langle \mu^2 \,, (\luno) \de_2 \phi  \rangle + \langle \mu^3 \,, (\luno) \de_3 \phi + \de_1 \phi \rangle\, . \nonumber
\end{split}
\ee
The three currents computes so far can be pulled back to the physical space $\mathcal{M}_\infty$.
As explained in the previous section the pull-back is done by taking the coordinate $\mu^k \,=\, 0$, and it coincides with the standard components $0jk$ of the generalized angular momentum of the Electromagnetic field.
Thus, we obtain that  the components of the momentum map $\mathbb{J}_{\mathcal{P}}$ corresponding to the subgroup $SO(3)$ of the Poincar\'e group $\mathscr{P}$ are exactly the three currents computed above
\be
{\mathbb{J}_{\mathcal{P}}}(\tilde{m}_\infty)_j \,=\, J_{\xi_{\mathscr{R}^{x^j}}} \;\;\; j \,=\, 1, 2, 3 \,.
\ee


\section*{Conclusions}
\addcontentsline{toc}{chapter}{Conclusions and discussion}

A global, covariant, geometric theory of symmetries has been constructed directly on the space of solutions of first order Hamiltonian field theories.  
The theory of symmetries of Hamiltonian systems, usually presented on an abstract setting on symplectic and pre-symplectic manifolds, is carried to the space of solutions of a first order Hamiltonian field theory which is naturally a pre-symplectic infinite-dimensional manifold.
The case of gauge theories, i.e., theories such that the space of solutions carries a presymplectic structure with non-trivial characteristic distribution was dealt by constructing a symplectic regularization of it using the coisotropic embedding theorem.

In particular, the case of geometrical symmetries emerging from transformations on the bundle and/or space-time supporting the theory were considered and we saw how, given a symmetry group of canonical transformations, it is possible to construct an algebra of conserved currents which is directly defined on the space of solutions of the theory.
We first considered the symplectic case where we analysed the Klein-Gordon field theory as a guiding example. 
Then, we considered the ``genuinely pre-symplectic'' case exemplified by gauge theories illustrated by Electrodynamics. 
Here we saw that a symmetry group may act both via dynamical and gauge symmetries and that the momentum map associated with such a symmetry group is reconstructed by the aid of a connection on the bundle associated with the gauge theory.

The reconstruction, by using an appropriate extension of the standard BRST theory, of the final manifold of constraints $\mathcal{M}_\infty$ of the theory, from the symmetries on the extended symplectic space $\mathcal{P}$ obtained by using the coisotropic embedding theorem will be discussed in further works.    In order to do so, the algebra of currents in $\mathcal{P}$ associated to a symmetry group will be studied further.

The case of symmetries of non-geometrical character, characteristic of hierarchies of integrable non-linear equations, will be discussed elsewhere.

\subsection*{Acknowledgments}

The authors acknowledge financial support from the Spanish Ministry of Economy and Competitiveness, through the Severo Ochoa Program for Centers of Excellence in RD (SEV-2015/0554).

The work has been supported by the Madrid Government (Comunidad de Madrid-Spain) under the Multiannual Agreement with UC3M in the line of “Research Funds for Beatriz Galindo Fellowships” (C\&QIG-BG-CM-UC3M), and in the context of the V PRICIT (Regional Programme of Research and Technological Innovation).

F. M. Ciaglia, A. Ibort, L. Schiavone would like to thank partial support provided by the research project QUITEMAD++, S2018/ TCS-A4342.

F. di Cosmo thanks the UC3M, the European Commission through the Marie Sklodowska-Curie COFUND Action (H2020-MSCA-COFUND-2017-GA 801538) and Banco Santander for their financial support through the CONEX-Plus Program.

A. Ibort and L. Schiavone would like to thank partial support provided by the MINECO research project PID2020-117477GB-I00.

G. Marmo would like to thank partial financial support provided by the Santander/UC3M Excellence Chair Program 2019/2020. He is also a member of the Gruppo Nazionale di Fisica Matematica (INDAM), Italy.

L. Schiavone would like to thank the support provided by Italian MIUR through the Ph.D. Fellowship at Dipartimento di Matematica R. Caccioppoli.


\bibliographystyle{alpha}

\end{document}